\documentclass{emulateapj}

\def\cm{\textrm{cm}}
\def\mum{\mu\textrm{m}}

\def\kpc{\textrm{kpc}}
\def\pc{\textrm{pc}}

\def\kms{\textrm{km}~\textrm{s}^{-1}}
\def\gcm2{\textrm{g}~\textrm{cm}^{-2}}
\def\ergscm3{\textrm{erg}~\textrm{s}^{-1}~\textrm{cm}^{-3}}
\def\ergcm3{\textrm{erg}~\textrm{cm}^{-3}}
\def\gscm2{\textrm{g}~\textrm{s}^{-1}~\textrm{cm}^{-2}}
\def\ergcmK34{\textrm{erg}~\textrm{cm}^{-3}~\textrm{K}^{-4}}
\def\cms31{\textrm{cm}^{-3}~\textrm{s}^{-1}}
\def\cmg21{\textrm{cm}^{2}~\textrm{g}^{-1}}

\def\phcm2s1{\textrm{photons}~\textrm{cm}^{-2}~\textrm{s}^{-1}}

\def\cm3{\textrm{cm}^{-3}}

\def\MHz{\textrm{MHz}}
\def\GHz{\textrm{GHz}}

\def\yr{\textrm{yr}}
\def\Myr{\textrm{Myr}}

\def\muGauss{\mu\textrm{G}}

\def\muJy{\mu\textrm{Jy}}
\def\nJy{\textrm{nJy}}
\def\mJy{\textrm{mJy}}

\def\GeVs1cm3{\textrm{GeV}~\textrm{s}^{-1}~\textrm{cm}^{3}}
\def\nWBackUnits{\textrm{nW}~\textrm{m}^{-2}~\textrm{sr}^{-1}}
\def\log{\textrm{log}}
\def\deg{\textrm{deg}}
\def\2phn{\phn\phn}
\def\3phn{\phn\phn\phn}
\def\4phn{\phn\phn\phn\phn}
\def\12phn{\4phn\4phn\4phn}
\def\Msun{\textrm{M}_{\sun}}
\newcommand{\mean}[1]{\ensuremath{\langle #1 \rangle}}

\newcommand{\Lsun}{\ensuremath{{\mathrm L}_{\odot}}}

\begin{document}

\title{The Physics of the FIR-Radio Correlation: II. Synchrotron Emission as a Star-Formation Tracer in High-Redshift Galaxies}
\author{Brian C. Lacki\altaffilmark{1} \& Todd A. Thompson\altaffilmark{1,2,3}}
\altaffiltext{1}{Department of Astronomy, The Ohio State University, 140 West 18th Avenue, Columbus, OH 43210, lacki@astronomy.ohio-state.edu}
\altaffiltext{2}{Center for Cosmology and Astroparticle Physics, The Ohio State University, 191 West Woodruff Avenue, Columbus, OH 43210.}
\altaffiltext{3}{Alfred P. Sloan Fellow}

\begin{abstract}
We construct one-zone steady-state models of cosmic ray (CR) injection, cooling, and escape over the entire dynamic range of the FIR-radio correlation (FRC), from normal galaxies to starbursts, over the redshift interval $0 \le z \le 10$.  Normal galaxies with low star-formation rates become radio-faint at high $z$, because Inverse Compton (IC) losses off the CMB cool CR electrons and positrons rapidly, suppressing their nonthermal radio emission.  However, we find that this effect occurs at higher redshifts than previously expected, because escape, bremsstrahlung, ionization, and starlight IC losses act to counter this effect and preserve the radio luminosity of galaxies.  The radio dimming of star-forming galaxies at high $z$ is not just a simple competition between magnetic field energy density and the CMB energy density; the CMB must also compete with every other loss process.  We predict relations for the critical redshift when radio emission is significantly suppressed compared to the $z \approx 0$ FRC as a function of star-formation rate per unit area.  For example, a Milky Way-like spiral becomes radio-faint at $z \approx 2$, while an M82-like starburst does not become radio-faint until $z \approx 10 - 20$.  We show that the ``buffering'' effect of non-synchrotron losses improves the detectability of star-forming galaxies in synchrotron radio emission with EVLA and SKA.
Additionally, we provide a quantitative explanation for the relative radio brightness of some high-$z$ submillimeter galaxies.  We show that at fixed star formation rate surface density, galaxies with larger CR scale heights are radio bright with respect to the FRC, because of weaker bremsstrahlung and ionization losses compared to compact starbursts.  We predict that these ``puffy starbursts'' should have steeper radio spectra than compact galaxies with the same star-formation rate surface density.  We find that radio bright submillimeter galaxies alone cannot explain the excess radio emission reported by ARCADE2, but they may significantly enhance the diffuse radio background with respect to a naive application of the $z \approx 0$ FRC.
\end{abstract}


\section{Introduction}
The radio and FIR luminosities of star-forming galaxies are linearly correlated over three decades in luminosity \citep[][]{vanDerKruit71,vanDerKruit73,deJong85,Helou85,Condon92,Yun01}.  This FIR-radio correlation (FRC) is driven by star-formation.  Starlight emitted by young, massive stars in the optical and UV is reprocessed by dust into far-infrared radiation.  At the same time, the supernova remnants of massive stars accelerate cosmic rays (CRs), which produce synchrotron radio emission in the galaxies' magnetic fields.  The relation at $z \approx 0$ is linear except in the very lowest luminosity galaxies \citep{Condon91b,Yun01,Bell03}.  Since the FRC has less than a factor of 2 scatter in the local Universe, radio luminosity is often used as a tracer of star-formation at both low and high redshift \citep[e.g.,][]{Cram98,Mobasher99}.

Although the FRC has been mainly studied in the low-$z$ universe, there has been recent interest in understanding how it applies to star-forming galaxies at high $z$.  There have been several conflicting results observationally.  A number of studies have found that the FRC holds unchanged or with little evolution at high redshifts (e.g., \citealt{Appleton04,Ibar08,Rieke09,Murphy09,Younger09,Garn09,Sargent10,Ivison10,Younger10}; \citealt{Persic07} suggest radio-dim ULIRGs at high $z$).  Submillimeter galaxies, however, seem to be radio-bright, by a factor of $\sim 3$ (\citealt{Kovacs06,Vlahakis07,Sajina08,Murphy09,Seymour09,Murphy09c,Michalowski09,Michalowski10}).  In particular, \citet{Murphy09} and \citet{Michalowski09} argue that their samples of submillimeter galaxies are intrinsically radio bright with respect to the local FRC, with no significant contamination from radio bright AGN.

In addition to the work on individual star-forming galaxies, there have also been new investigations into both the recently-resolved FIR and the unresolved GHz radio backgrounds.  The FIR background was detected by COBE and has long been attributed to star-formation \citep[see the reviews by][]{Hauser01,Lagache05}.  The star-formation origin of the FIR background was recently confirmed by BLAST \citep{Devlin09,Pascale09}.  A detection of the extragalactic GHz radio background has also been recently reported by ARCADE2 \citep{Fixsen09,Seiffert09}.  Predictions for the strength of the diffuse radio background from star formation rely on the FRC holding out to $z \approx 1 - 2$ \citep[e.g.,][]{Haarsma98,Dwek02}, where most star-formation occurs.

In this paper, we describe how and why the linearity and normalization of the FRC are broken (or preserved) for different types of galaxies at high $z$.  We describe the basic physics of our model and the elements of the high-$z$ universe that physically allow the FRC to break down in \S~\ref{sec:Theory}.  In \S~\ref{sec:Procedure}, we briefly describe our underlying procedure.  In \S~\ref{sec:Results}, we discuss our results, providing predictions for when star-forming galaxies should be more or less radio luminous than the $z \approx 0$ FRC suggests.   We also summarize the implications of our predictions for radio emission as a star formation tracer in \S~\ref{sec:SFTracer} and estimate the effects of radio-bright SMGs on the diffuse radio background in \S~\ref{sec:RadioBackground}.

Throughout this paper, primed quantities denote the source's rest-frame and unprimed quantities are for the observer-frame.  

\section{Theory}
\label{sec:Theory}

\citet{Lacki09} (LTQ) recently presented a theory of the FRC over its entire span at low redshift, from normal galaxies like the Milky Way to the densest ULIRGs like Arp 220.  

The linearity and normalization of the FRC provide strong constraints on the magnetic energy density in galaxies, which governs the strength of synchrotron losses.  Indeed, the energy density of starlight in dense galaxies is so large that Inverse Compton losses alone shorten the CR electron and positron lifetime to $\sim 10^4~\yr$ in galaxies like Arp 220 \citep{Condon91}.  The magnetic field in Arp 220-like starbursts must exceed mG for the cosmic ray electrons and positrons to radiate a large enough fraction of their energy in synchrotron emission to account for the observed GHz radio emission.  Such galaxies are likewise sufficiently dense that ionization and bremsstrahlung losses are also rapid.  These considerations imply that the magnetic energy density must increase dramatically from normal spirals to dense bright starbursts.  The observed Schmidt law allows us to quantify these dependences since it connects the star-formation rate surface density ($\Sigma_{\rm SFR}$), which is proportional to the energy density in starlight ($U_{\star} = F_{\star} / c \propto \Sigma_{\rm SFR}$), to the gas surface density: $\Sigma_{\rm SFR} \propto \Sigma_g^{1.4}$ according to \citet{Kennicutt98}, or $\Sigma_{\rm SFR} \propto \Sigma_g^{1.7}$ according to \citet{Bouche07}.  Using the \citet{Kennicutt98} Schmidt Law, LTQ found that the magnetic field strength $B$ must increase as $B \propto \Sigma_g^{0.6 - 0.8}$ or $B \propto \rho^{0.5 - 0.6}$, where $\Sigma_g = 2 \rho h$, implying almost a thousandfold increase in $B$ from Milky Way-like galaxies to Arp 220-like ULIRGs.  This strong dependence of $B$ on the global properties of the galaxy is largely responsible for the radio emission from starbursts and is required by the existence of the FRC.  It is also crucial to the physics of how the FRC is maintained or broken at high z.  LTQ was unable to distinguish between the possibilities of $B$ scaling with $\Sigma_g$ and $B$ scaling with $\rho$.

LTQ found that over most of the range of the FRC, normal galaxies and starbursts are calorimetric.  They are electron calorimeters, meaning CR electrons lose most of their energy before escaping.  They are also UV calorimeters, meaning UV light is efficiently absorbed by dust and converted to FIR light.  If CR electrons cool mainly by synchrotron losses, then the ratio of FIR emission to radio emission from primary CR electrons should be constant, since each is some fraction of the power released by star-formation.  This is the essence of calorimeter theory, first proposed by \citet{Volk89}.  However, LTQ found that even in this strict calorimeter limit, $L_{\rm FIR}/L_{\rm radio}$ should decrease systematically by $\sim 2$ from low surface brightness galaxies to dense ULIRGs.  The critical synchrotron frequency causes this ``$\nu_C$ effect'': at fixed frequency, as the magnetic field strength increases with increasing surface density, lower energy electrons dominate the synchrotron emission because the synchrotron frequency of electrons, $\nu_C$, is proportional to $E_e^2 B$, where $E_e$ is the electron energy.  Since the injected CR spectrum is likely steeper than $E_e^{-2}$, there is more power radiated at low energies.  Thus, radio emission increases with surface density of gas and star formation, even including only synchrotron losses in the pure electron and UV calorimeter limit.

On top of calorimetry, LTQ proposed that two different conspiracies combine to produce the linear FRC, each working in a different surface density regime.  The ``low-$\Sigma_g$ conspiracy'' arises as both UV calorimetry and electron calorimetry gradually fail at relatively low surface densities\footnote{$1~\gcm2 = 4800~\Msun \pc^{-2}$.} ($\Sigma_g < 0.01~\gcm2$; $\Sigma_{\rm SFR} \la 0.06~\Msun~\kpc^{-2}~\yr^{-1}$).  Cosmic ray electrons escape diffusively without radiating all of their energy, decreasing the radio emission from the calorimetric expectation.  However, some UV photons also escape without being reprocessed into FIR by absorption.  Quantitatively, each effect alters the luminosity by a factor of $\sim 5$ in the very lowest surface density galaxies ($\Sigma_g \approx 10^{-3}~\gcm2$; $\Sigma_{\rm SFR} \approx 0.001 - 0.002~\Msun \kpc^{-2} \yr^{-1}$).  Therefore, both the FIR and radio emission decrease roughly equally, to maintain a constant $L_{\rm FIR}/L_{\rm radio}$ \citep[see also][]{Bell03}.

At the other extreme, high surface density compact starbursts have two completely different opposing processes that affect the radio emission.  Compact starbursts are expected to be proton calorimeters: CR protons lose almost all of their energy through pion interactions with the dense ISM of starbursts, emitting $\gamma$-rays, neutrinos, and secondary electrons and positrons \citep[][LTQ]{Rengarajan05,Loeb06,Thompson07}.  The secondary electrons and positrons themselves emit synchrotron radio emission, and outnumber the primary CR electrons.  Proton calorimetry therefore greatly enhances the radio emission, all else being equal.  Although escape losses are negligible in starbursts, synchrotron does compete with the other important cooling processes: Inverse Compton, bremsstrahlung, and ionization \citep[][LTQ]{Thompson06}.  These are particularly important in bright, dense starbursts, because they have very high gas densities and photon energy densities.  These other losses reduce the share of CR electron and positron power going into radio by a factor of $\sim 10 - 20$.  Between the secondaries and the $\nu_C$ effect increasing $L_{\rm radio}$, and the non-synchrotron losses suppressing $L_{\rm radio}$, a linear FRC correlation is produced.  This is the ``high-$\Sigma_g$ conspiracy'' of LTQ, and it operates to produce a linear FRC even though these starbursts are completely calorimetric (for protons, electrons, and positrons) and escape losses are negligible.

Galaxies at high redshift can be used to test the theory of the FRC presented in LTQ.  At high $z$, we expect that both the low- and high-$\Sigma_g$ conspiracies are unbalanced for some galaxies.  At low $\Sigma_g$, the energy density of the CMB ($U_{\rm CMB}$) competes with the energy density in both starlight and magnetic fields.  CRs in galaxies face increasing losses from Inverse Compton since the CMB is stronger.  Normal galaxies therefore should be radio dim, as Inverse Compton losses divert power from synchrotron.  Starbursts, however, where CRs already face intense IC losses from starlight, should not be affected except at the highest redshifts \citep[e.g.,][]{Condon92,Carilli99,Carilli08}.  In this paper, we quantify how much the radio emission is dimmed by increased IC losses from the CMB as a function of $z$.

On the other hand, high-$z$ starbursts often have different morphologies than those observed of low-$z$ ULIRGs.  Local starbursts including ULIRGs are typically compact, with radii of a few hundred parsecs and vertical scales heights of less than a hundred parsecs \citep{Solomon97,Downes98}.  These properties do not hold for submillimeter galaxies (SMGs), very bright ($L \ga 10^{12} \Lsun$) starbursts at $z \approx 2$ mostly powered by star-formation and not AGNs \citep[e.g.,][]{Pope06,Valiante07,Watabe09}.  SMGs are usually several kpc in diameter (several times larger than low-$z$ ULIRGs) \citep[e.g.,][]{Chapman04,Biggs08,Younger08,Iono09,Younger10}, and have much lower surface densities than their low-$z$ counterparts of the same luminosity (\citealt{Tacconi06}, but see \citealt{Walter09}).  Submillimeter galaxies have large random velocities compared to their rotation speeds, implying scale heights of $h \approx 1~\kpc$ \citep{Tacconi06,Genzel08,Law09}.  At high redshift, we therefore must consider a class of \emph{puffy starbursts}\footnote{We use the term ``puffy'' instead of ``extended'' to emphasize that the gas and star-formation is extended \emph{vertically} as well as radially.}, with $\Sigma_g \ge 0.1~\gcm2$ ($\Sigma_{\rm SFR} \ga 2 - 4~\Msun \kpc^{-2} \yr^{-1}$) but scale heights of $h = 1~\kpc$, in addition to the $h = 100~\pc$ ``compact starbursts'' typical of local ULIRGs considered in LTQ.  

Importantly, puffy starbursts will have a smaller volume density for a given $\Sigma_g$ and $\Sigma_{\rm SFR}$.  Since bremsstrahlung, ionization, pion, and possibly synchrotron losses all depend on the volume density instead of surface density, these loss processes may all be weaker in puffy starbursts like SMGs.  With these losses suppressed, the high-$\Sigma_g$ conspiracy will become unbalanced, and the FRC can be broken.  These starbursts can either be radio-bright or radio-dim, depending on their magnetic field strengths.  Since the scale height is so large and the volume density is relatively low for SMGs, we can determine whether $B$ scales with $\Sigma_g$ or $\rho$ with these galaxies, breaking the degeneracy in LTQ.  Because $\Sigma_g = 2 \rho h$, if $B$ increases with $\rho$, then the magnetic field in puffy starbursts will be weak and synchrotron radio emission will be dim compared to compact starbursts with the same $\Sigma_g$.  However, if $B$ increases with $\Sigma_g$, the magnetic field will be strong and synchrotron radio emission will be bright in puffy starbursts compared to compact starbursts with the same $\Sigma_g$.

\section{Procedure and Assumptions}
\label{sec:Procedure}
We model galaxies and starbursts as uniform disks of gas, with scale height $h$, star formation rate surface density\footnote{In our models, the rest-frame bolometric luminosity from star-formation $F^{\prime}$ is directly proportional to $\Sigma_{\rm SFR}$.  The conversion factor is $F^{\prime} = 5.6 \times 10^9 \Sigma_{\rm SFR,0} \Lsun~\kpc^{-2}$, where $\Sigma_{\rm SFR,0}$ is the star formation rate surface density in units of $\Msun~\kpc^{-2}~\yr^{-1}$.} $\Sigma_{\rm SFR}$, and gas surface density $\Sigma_g$.  We solve the diffusion-loss equation to find the steady-state equilibrium CR spectra in galaxies and starbursts.  Injection, escape, and cooling losses all compete at each energy to determine the final CR spectrum (see LTQ for details).  

The relevant scale height is the height of the volume in which CRs are confined and produce synchrotron radiation.  Normal galaxies ($\Sigma_g \le 0.01~\gcm2$ or $\Sigma_{\rm SFR} \la 0.06~\Msun \kpc^{-2} \yr^{-1}$) have large radio halos with $h \approx 1~\kpc$, which we adopt as the CR scale height, even though the gas disk is much thinner.  Compact starbursts are much smaller, with $h \approx 100~\pc$ \citep[e.g.,][]{Solomon97,Downes98}; we use this as their scale height because they are probably good enough calorimeters to prevent most of the CR electrons/positrons from escaping out into the galactic haloes at high enough surface densities ($\Sigma_g \ga 0.1~\gcm2$ or $\Sigma_{\rm SFR} \ga 2 - 4~\Msun \kpc^{-2} \yr^{-1}$).  The CR disk scale height for our prototypical puffy starbursts, the SMGs, is not yet directly measured, but it must be at least the gas scale height, if the star formation is distributed throughout the gas.  The gas scale height for SMGs is kinematically inferred to be $h \approx 1~\kpc$ \citep[e.g.,][]{Tacconi06,Genzel08,Law09}, which we adopt as the CR scale height of puffy starbursts.

Our models span the entire observed range in $\Sigma_g$ for galaxies and starbursts, from $0.001~\gcm2$ to $10~\gcm2$.  Combined with the scale height, $\Sigma_g$ determines the rate of injection, escape, and cooling at all energies.  The total rate of injection per unit volume is proportional to the volumetric star-formation rate, $\Sigma_{\rm SFR} / (2h)$.  The observed Schmidt law directly connects $\Sigma_g$ and $\Sigma_{\rm SFR}$ \citep{Schmidt59}.  We consider both $\Sigma_{\rm SFR} \propto \Sigma_g^{1.4}$ \citep[][hereafter K98]{Kennicutt98} and $\Sigma_{\rm SFR} \propto \Sigma_g^{1.7}$ \citep[][hereafter B07]{Bouche07}, using the normalizations given by K98 and B07 respectively.  The B07 relation was explicitly derived for high-$z$ galaxies, including SMGs.  Protons and electrons are injected into our model galaxies with an energy spectrum $E^{-p}$; in this paper, we use $p = 2.2$ in almost all cases.\footnote{The exception is the ``Standard Model'' from LTQ, which assumes $p = 2.3$.  The slightly different $p$ makes small quantitative differences, but does not change our conclusions.}

Escape, synchrotron, Inverse Compton, bremsstrahlung, and ionization losses are included for electrons and positrons, along with escape, ionization, and pion losses for protons.  Diffusive escape times are normalized to the Milky Way value, with a loss time of $t_{\rm diff} \propto E^{-1/2}$.  We also consider variants with advective escape, as in starburst super-winds.  We assume that cosmic rays travel through gas with density $f\mean{n}$, where $\mean{n}$ is the mean ISM number density and $1.0 \le f \le 2.0$.  The magnetic field is parametrized as $B \propto \Sigma_g^a$ or $B \propto \rho^a$, normalized by the Milky Way magnetic field strength.  We searched for models that satisfy the local $z \approx 0$ FRC for normal galaxies and compact starbursts.  The CR proton spectrum is then normalized by Milky Way CR proton constraints.  The parameters chosen for each variant are used to predict the FRC for puffy starbursts and at high $z$.

In this paper, we consider several variants to get a sense of how the FRC varies with redshift:
\begin{itemize}
\item The LTQ standard model, with no winds, $B \propto \Sigma_g^a$, and the K98 star-formation relation.

\item The LTQ model with winds of $300~\kms$ in starbursts ($\Sigma_g \ge 0.1~\gcm2$), $B \propto \rho^a$, and the K98 star-formation relation.

\item A variant with no winds, $B \propto \Sigma_g^a$, and the B07 star-formation relation.

\item A variant with winds of $300~\kms$ in starbursts ($\Sigma_g \ge 0.1~\gcm2$), $B \propto \Sigma_g^a$, and the B07 star-formation relation.

\item A variant with winds of $300~\kms$ in starbursts ($\Sigma_g \ge 0.1~\gcm2$), $B \propto \rho^a$, and the B07 star-formation relation.
\end{itemize}
We show in Table~\ref{table:Models} that we can reproduce an acceptably linear local FRC for each variant.  We are able to adopt parameters consistent with Milky Way-derived CR proton constraints such that $L_{\rm TIR}/L_{\rm radio}$ varies by a factor of 1.7 - 2.2 from normal galaxies to compact starbursts.  This variation is consistent with the factor of $\sim 2$ scatter in the FRC\footnote{The biggest exception is the variant with $B \propto \Sigma_g^a$ and winds, in which the winds invariably make electron escape too efficient when $\Sigma_g = 0.1~\gcm2$ ($\Sigma_{\rm SFR} \approx 4~\Msun~\kpc^{-2}~\yr^{-1}$ with B07), suppressing its radio luminosity (see the detailed discussion in LTQ, Appendix A.2).  We are unable to find any parameters where the variation in $L_{\rm TIR}^{\prime}/L_{\rm radio}^{\prime}$ is less than 2.0, but we consider a variation of 2.2 adequate for this paper's purposes. \label{ftnt:WindVariant}} \citep[e.g.,][]{Yun01}.  As in LTQ, the need for a linear local FRC strongly constrains the magnetic field in galaxies to scale as either $\Sigma_g^{0.7 - 0.8}$ or $\rho^{0.5 - 0.6}$ (see \S~\ref{sec:Theory}).

The dependence of the FIR emission at a given wavelength is beyond the scope of this paper, since it depends on the exact SED of the FIR emission in the galaxy.  Instead, we simply use the Total Infrared (TIR) emission, assuming it is all of the UV light reprocessed by the dust, and use $L_{\rm TIR}^{\prime}$ as a proxy for $L_{\rm FIR}^{\prime}$.  \citet{Calzetti00} estimate that $L_{\rm TIR}^{\prime} \approx 1.75 L_{\rm FIR}^{\prime}$, a correction we apply to the local value of $L_{\rm FIR}^{\prime}/L_{\rm radio}^{\prime}$ from \citet{Yun01} to find the normalization of $L_{\rm TIR}^{\prime}/L_{\rm radio}^{\prime}$ at $z \approx 0$ (see LTQ).\footnote{When calculating the $q_{\rm FIR}^{\prime}$ observable defined in \citet{Helou85}, we then divide our calculated $L_{\rm FIR}^{\prime}$ by $L_{\rm TIR}^{\prime}/L_{\rm FIR}^{\prime}$ to get the true FIR luminosity.  However, we do not account for different $L_{\rm TIR}^{\prime}/L_{\rm FIR}^{\prime}$ values for normal galaxies or SMGs.}  We assume that the total TIR emissivity (defined here as luminosity per volume) has been measured and corrected for redshift to its rest-frame value, $\epsilon_{\rm TIR}^{\prime}$.  We can also calculate the observable quantity $q_{\rm FIR}$ as $\log_{10}(L_{\rm TIR}/L_{\rm radio}) - 3.67$ \citep{Helou85}.

For the radio emission, we calculate both the rest-frame radio emissivity $\epsilon_{\rm radio, rest}^{\prime} = \nu^{\prime} \epsilon_{\nu}^{\prime} (\nu^{\prime})$ where $\nu^{\prime} = 1.4~\GHz$, and the rest-frame radio emissivity as estimated by an observer using a k-correction.  Note that we are \emph{not} actually trying to calculate the specific flux in the observer frame, but instead what an observer will infer for the \emph{rest-frame} flux after using a k-correction\footnote{The situation is essentially the same as an observer measuring the radio flux of a $z = 0$ galaxy at 2 GHz and trying to infer the 1 GHz flux by using the spectral slope.  The observer has an observed radio flux at 1.4 GHz, and can easily calculate the rest-frame flux at 1.4 (1 + z) GHz flux, but wants the rest-frame flux at 1.4 GHz.}: the rest-frame flux is what matters when we are considering the true, intrinsic evolution of the FRC.  Hence, although the specific flux in the observer-frame will have a bandwidth compression factor, we assume the observer will take it back out to get the inferred rest-frame specific flux.  To calculate the observer-inferred rest-frame emissivity at rest-frame frequency $\nu^{\prime} = 1.4~\GHz$, we start with $\nu^{\prime} \epsilon_{\nu}^{\prime}(\nu^{\prime}_{\rm obs})$, where $\nu^{\prime}_{\rm obs} = (1 + z) \nu$ and $\nu = \nu^{\prime} = 1.4~\GHz$.  Since the synchrotron spectra are typically expected to fall off as $\epsilon_{\nu}^{\prime} \propto \nu^{\prime -0.7}$ from observations of local star-forming galaxies, the radio luminosity can be k-corrected by multiplying by $(1 + z)^{0.7}$.  We therefore calculate $\epsilon_{\rm radio, inf}^{\prime} = \nu^{\prime} \epsilon_{\nu}^{\prime}(\nu^{\prime}_{\rm obs}) (1 + z)^{0.7}$, which we will refer to as the ``inferred'' radio emissivity.  The ratio of the TIR and radio luminosities $L_{\rm TIR}^{\prime} / L_{\rm radio}^{\prime}$ is then $\epsilon_{\rm TIR}^{\prime} / \epsilon_{\rm radio,rest}^{\prime}$ in the rest frame, and would be inferred to be $\epsilon_{\rm TIR}^{\prime} / \epsilon_{\rm radio, inf}^{\prime}$ in the rest-frame by observers.

\section{Results \& Discussion}
\label{sec:Results}
\label{sec:Discussion}

\subsection{The Evolving FIR-Radio Correlation}
\label{sec:FRCEvolution}

\begin{figure*}
\centerline{\includegraphics[width=9cm]{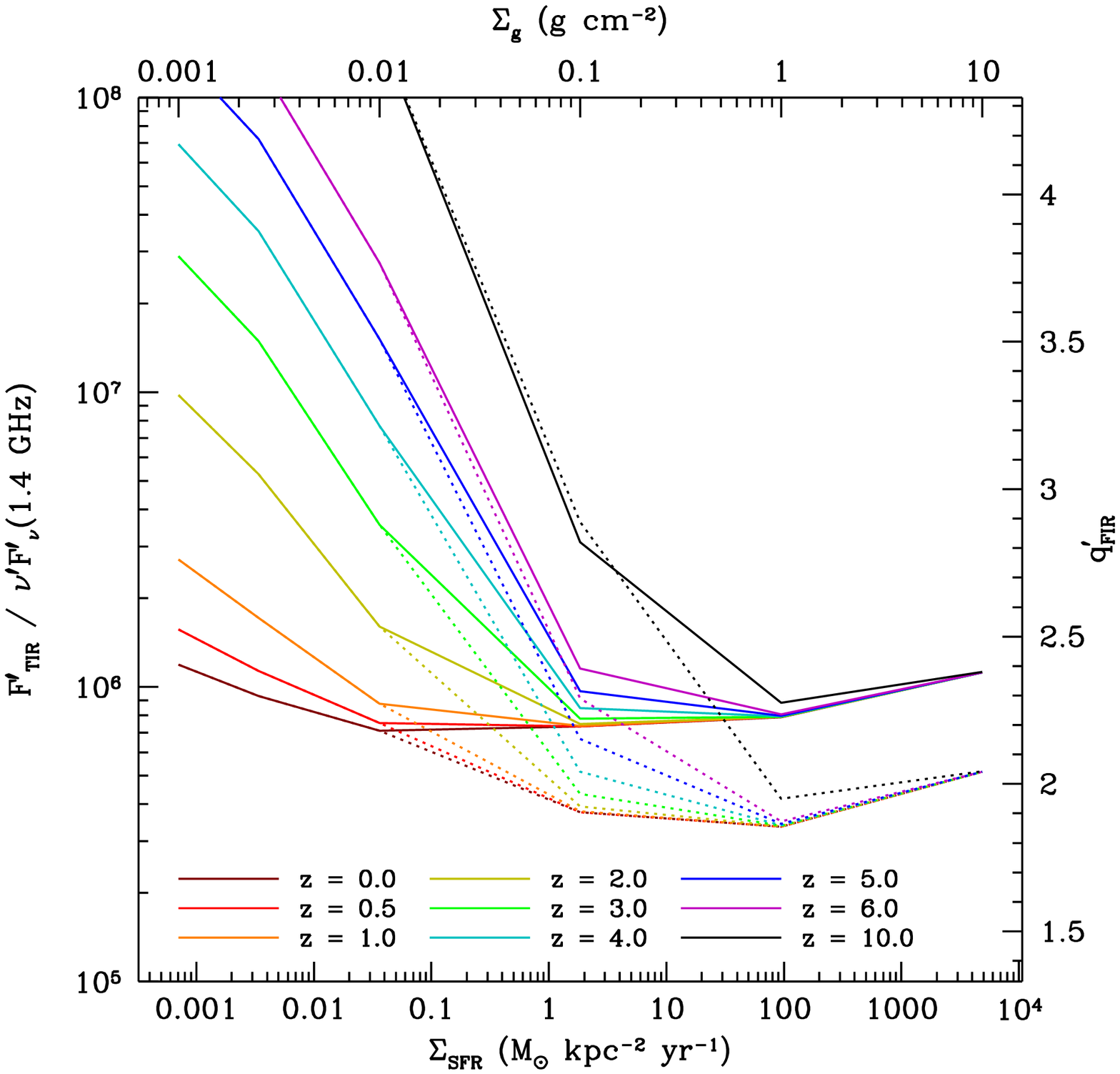}\includegraphics[width=9cm]{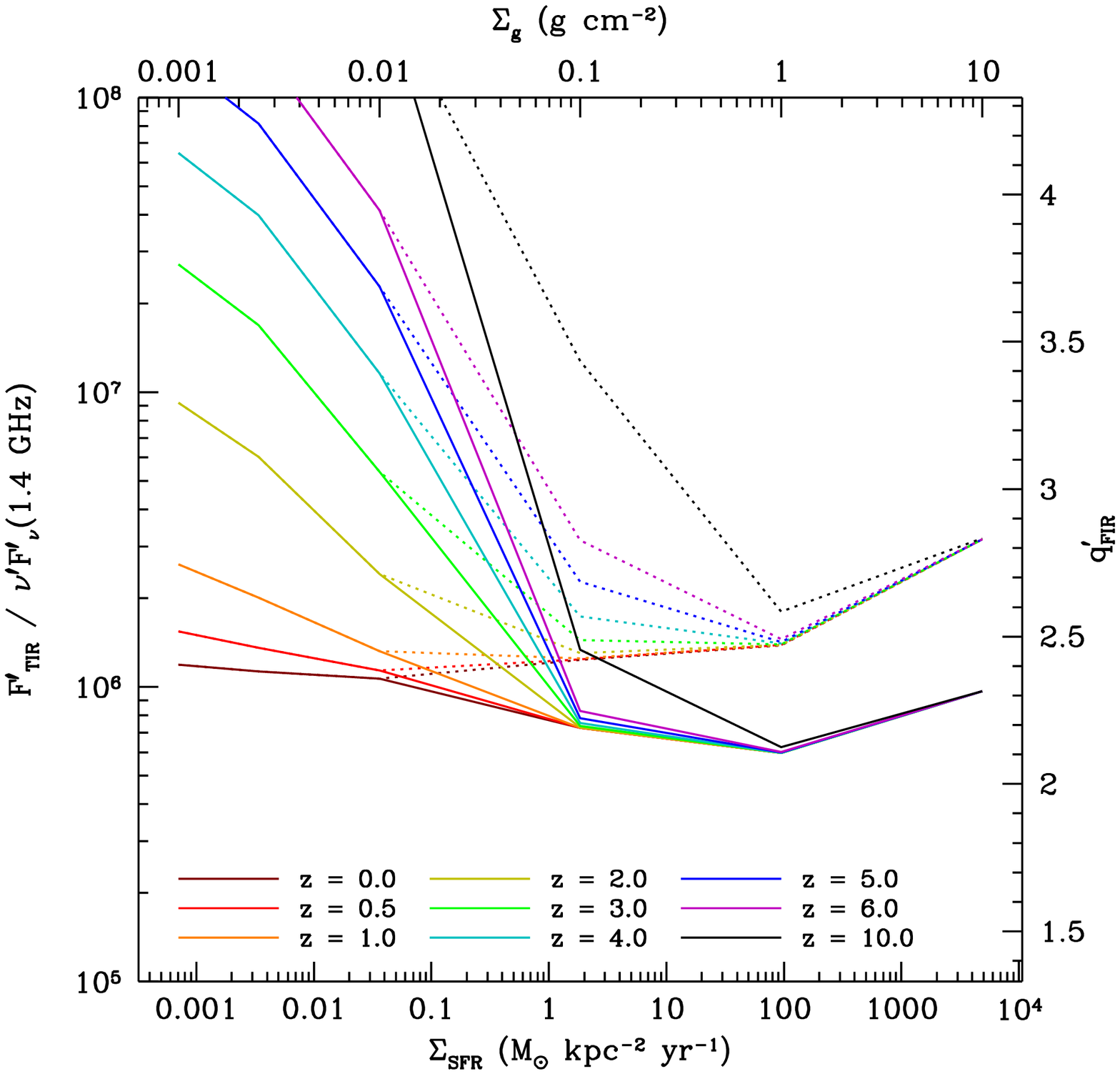}}
\figcaption[simple]{The FIR-radio correlation at high redshift, in the rest-frame, using the B07 star formation law.  On left is the case with $B \propto \Sigma_g^{0.7}$ and no winds; on right, the case with $B \propto \rho^{0.6}$ and winds.  Solid lines have $h = 100~\pc$ for starbursts ($\Sigma_g \ge 0.1~\gcm2$; $\Sigma_{\rm SFR} \ge 2 - 4~\Msun~\kpc^{-2}~\yr^{-1}$), while dotted lines have $h = 1~\kpc$ for starbursts.  Both panels show that a linear FRC is produced at $z = 0$ with the correct normalization.  Galaxies become radio-dim at high redshift as IC losses off the CMB increase.  Note that in each case the puffy starbursts do not lie on the same FRC as compact starbursts: there is a systematic offset caused by the unbalancing of the high-$\Sigma_g$ conspiracy (\S~\ref{sec:Puffy}).  We do not include thermal radio emission, which will set a minimum radio luminosity or maximum $q^{\prime}_{\rm FIR}$ at each $\Sigma_g$.\label{fig:LFIRRadioRest}}
\end{figure*}

\begin{figure*}
\centerline{\includegraphics[width=9cm]{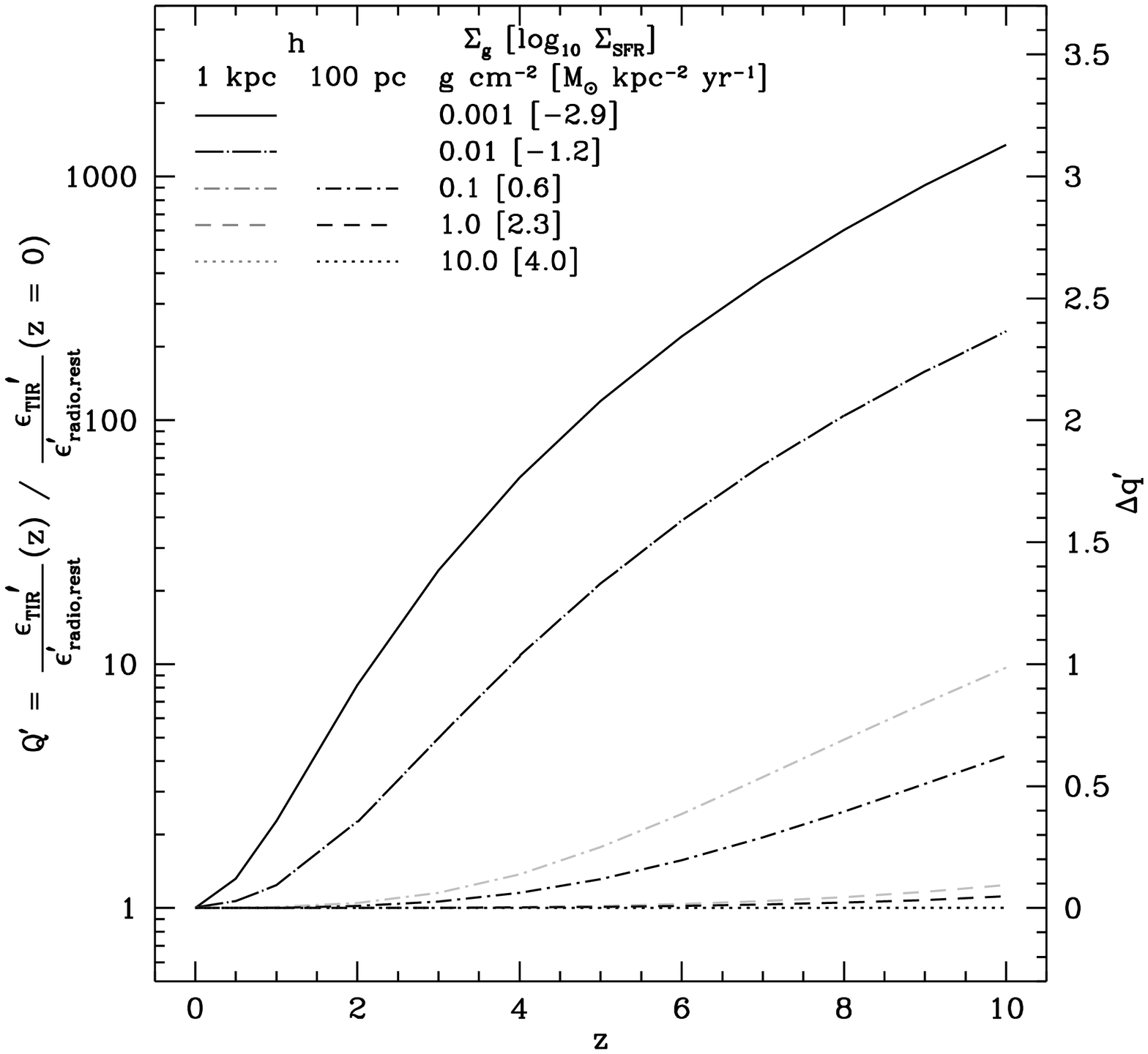}\includegraphics[width=9cm]{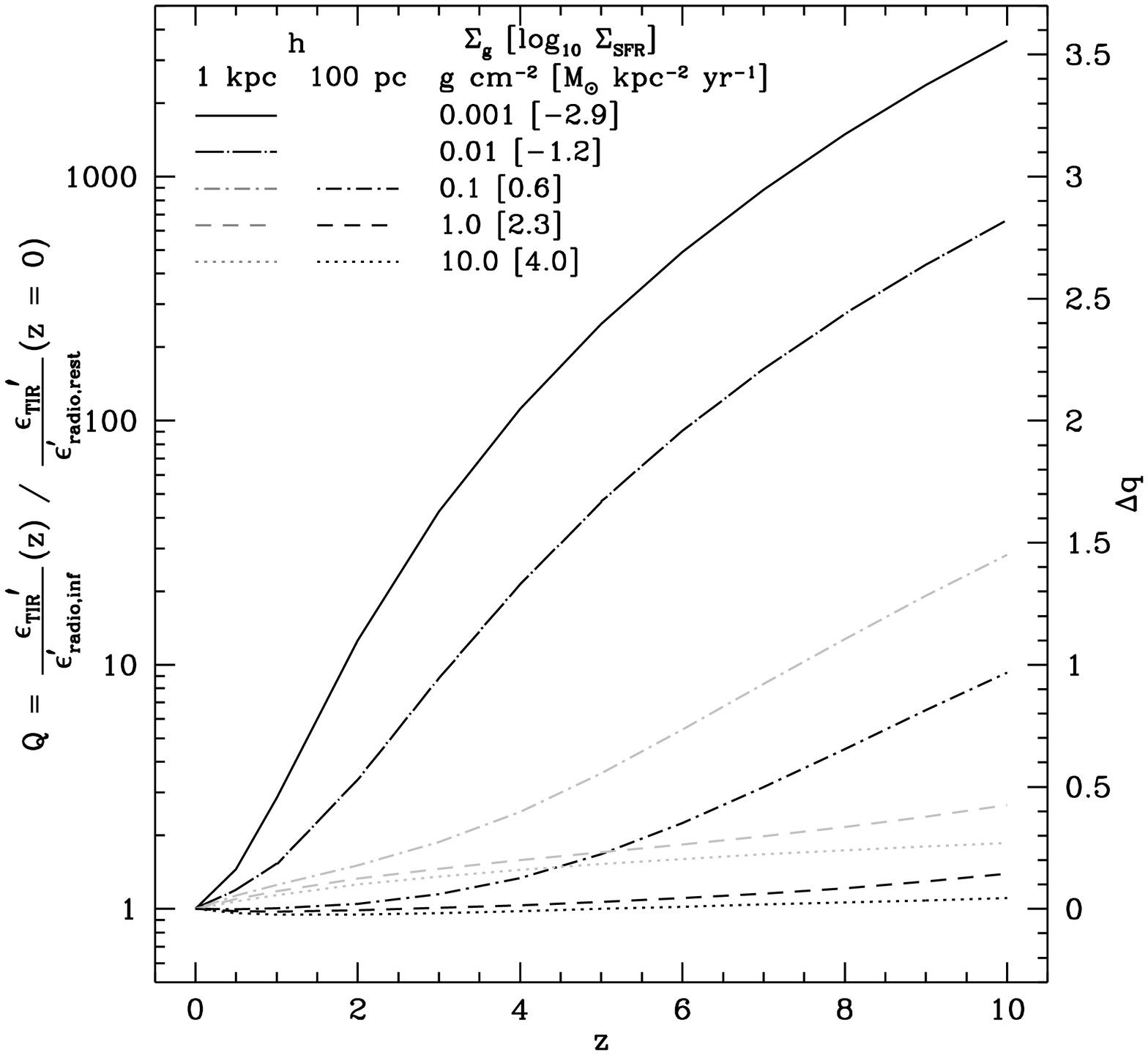}}
\figcaption[simple]{The evolution of the FIR-radio correlation, both in the rest-frame at $\nu^{\prime} = 1.4~\GHz$ (\emph{left panel}), and as inferred for the rest-frame from observations at $\nu = 1.4~\GHz$ (\emph{right panel}).  Black lines have $h = 100~\pc$ for starbursts, while grey lines have $h = 1~\kpc$ for starbursts.  Both panels show the case with $B \propto \Sigma_g^{0.7}$, no winds, and the B07 star-formation law.  Normal galaxies become radio dim because of IC losses off the CMB at intermediate redshift.  Both compact and puffy starbursts maintain their rest-frame radio luminosities until high redshift.  All galaxies are ``buffered'' by non-synchrotron losses (\S~\ref{sec:FRCEvolution} and Appendix~\ref{sec:CMBRadioDim}), so that the evolution is not as great as would be expected with only synchrotron and IC losses off the CMB.  The inferred evolution is usually greater than the rest-frame evolution, because normal galaxies and puffy starbursts have $\alpha \ga 0.7$. \label{fig:FRCEvolution}}
\end{figure*}

We show the rest-frame FRC for our model with $B \propto \Sigma_g^{0.7}$, the B07 star-formation law, and no winds, in Figure~\ref{fig:LFIRRadioRest} (\emph{left panel}) as an example.  The solid dark red line is the $z = 0$ FRC.  All solid lines assume that starbursts are compact, with $h = 100~\pc$.  It is clear from Figure~\ref{fig:LFIRRadioRest} that compact starbursts show little evolution in the FRC, while low surface density galaxies have lower radio luminosities at high redshift.  This behavior is robust in all of the variants.  The cause of evolution in the FRC is Inverse Compton losses off the CMB for CR electrons and positrons.  Figure~\ref{fig:FRCEvolution} (\emph{left panel}) shows that the FRC should display relatively little evolution out to $z \approx 1$, except for the lowest surface brightness galaxies.  However, normal galaxies have suppressed synchrotron radio emission at $z \approx 2$, a factor of $\sim 2$ for $\Sigma_g = 0.01~\gcm2$ ($\Sigma_{\rm SFR} \approx 0.06~\Msun~\kpc^{-2}~\yr^{-1}$) and of order 10 for $\Sigma_g = 0.001~\gcm2$ ($\Sigma_{\rm SFR} \approx 0.001 - 0.002~\Msun~\kpc^{-2}~\yr^{-1}$).  The radio luminosities continue to fall with redshift.  At $z \approx 5$, IC off the CMB starts to matter even for the weaker starbursts ($\Sigma_g = 0.1~\gcm2$; $\Sigma_{\rm SFR} \approx 2 - 4~\Msun~\kpc^{-2}~\yr^{-1}$).  Dense starbursts ($\Sigma_g = 10~\gcm2$; $\Sigma_{\rm SFR} \approx 900~\Msun~\kpc^{-2}~\yr^{-1}$ for the K98 law or $9000~\Msun~\kpc^{-2}~\yr^{-1}$ for the B07 law) remain on a linear FRC even at $z \approx 10$.

The strong cooling from bremsstrahlung, ionization, and IC off starlight implied by the high-$\Sigma_g$ conspiracy (\S~\ref{sec:Theory}) acts as a buffer against IC losses off the CMB.  Similarly, diffusive escape provides a similar buffering effect in low-$\Sigma_g$ galaxies, so that the increased IC losses off the CMB does not suppress the radio emission quickly.  In other words, most of the power from IC scattering of CMB photons does not come at the expense of synchrotron, but of other losses.  The radio dimming cannot be described by a simple competition between magnetic field energy density and the CMB energy density; the CMB energy density must also compete with \emph{every other loss process}.  The buffering actually serves as an important test for both conspiracies described in \S~\ref{sec:Theory}.  In high-$\Sigma_g$ galaxies and starbursts, we predict that bremsstrahlung, ionization, and Inverse Compton of starlight already take a large portion of a \GHz~electron's energy budget; hence, the high-$\Sigma_g$ conspiracy works to hold $L_{\rm TIR}^{\prime}/L_{\rm radio}^{\prime}$ constant out to quite high redshift.  The buffering essentially doubles the redshift that compact starbursts remain on the FRC; the weakest starbursts ($\Sigma_g = 0.1~\gcm2$; $\Sigma_{\rm SFR} \approx 2 - 4~\Msun~\kpc^{-2}~\yr^{-1}$) are radio-dim by a factor of $3$ at $z = 10$ with buffering, instead of $z = 5$ without the buffering.  In low-$\Sigma_g$ galaxies, electrons and positrons can easily escape before they lose energy to Inverse Compton off the CMB at low enough $z$; thus, the low-$\Sigma_g$ conspiracy also works to hold $L_{\rm TIR}^{\prime}/L_{\rm radio}^{\prime}$ constant.

The suppression due to IC losses off the CMB seen in Figure~\ref{fig:LFIRRadioRest} can be estimated by examining the ratios of the synchrotron cooling time to the total loss time, including both escape and cooling losses.  In Appendix~\ref{sec:CMBRadioDim}, we derive the ratios of loss times and we show that for a given Schmidt law, a critical redshift $z_{\rm crit}$ can be defined for each $\Sigma_g$ and $\Sigma_{\rm SFR}$ at which radio emission is suppressed.  We define $z_{\rm crit}$ to be the redshift when the rest-frame radio luminosity at $\nu^{\prime} = 1.4~\GHz$ is suppressed by a factor of 3 compared to $z = 0$.  For our model with the B07 star-formation law, no winds, and $B \propto \Sigma_g^{0.7}$, we find the critical redshifts are approximated as 
\begin{equation}
\label{eqn:B07zCrit}
z_{\rm crit} \approx \left\{ \begin{array}{ll}
			1.4 & ({\rm Normal~galaxies}, \Sigma_{\rm SFR} \la 0.02)\\
			5.8 \Sigma_{\rm SFR}^{0.23} - 1 & ({\rm Normal~galaxies}, \Sigma_{\rm SFR} \ga 0.02)\\
			5.7 \Sigma_{\rm SFR}^{0.23} - 1 & ({\rm Puffy~starbursts})\\
			7.4 \Sigma_{\rm SFR}^{0.23} - 1 & ({\rm Compact~starbursts}),
			\end{array} \right.
\end{equation}
where $\Sigma_{\rm SFR}$ is in units of $\Msun~\kpc^{-2}~\yr^{-1}$.  We refer the reader to Appendix~\ref{sec:CMBRadioDim}, where we present similar relations for our other models of the FRC.

We note that the radio suppression could, conceivably, be used to measure the temperature of the CMB at high redshift.  In principle, this method could apply to any galaxy with a radio and FIR detection.  However, the conspiracies would have to be accounted for, and they are affected significantly by both the gas surface density $\Sigma_g$ and scale height $h$ (see \S~\ref{sec:Puffy}).  Any measurement of the CMB temperature would depend on assumptions of the galaxy properties and would be model-dependent.  

As Figure~\ref{fig:FRCEvolution} (\emph{right panel}) shows, the inferred rest-frame values of $L^{\prime}_{\rm TIR}/L^{\prime}_{\rm radio}$ show additional apparent evolution, simply because the spectral slopes of the galaxies are not exactly $0.7$.  Compact starbursts have flatter spectra\footnote{In this paper (unlike LTQ), $\alpha^{\prime} \equiv \frac{d \log F^{\prime}_{\nu}}{d \log \nu^{\prime}}$ and $\alpha \equiv \frac{d \log F_{\nu}}{d \log \nu}$ are instantaneous spectral slopes, not the measured spectral slopes between two observed frequencies, unless otherwise noted. \label{ftnt:AlphaNotation}} with $\alpha \approx 0.4 - 0.6$ at 1.4 GHz, so by adopting $\alpha = 0.7$ they \emph{appear} to become slightly radio brighter until $z \approx 1$, after which at higher frequencies their spectra steepen, and their radio emission appears to dim again at higher $z$.  Normal galaxies have steeper spectra with $\alpha \approx 0.9 - 1.0$ at 1.4 GHz, so their apparent radio luminosity sinks below their true radio luminosity at high redshift.

At higher redshifts, our models predict that galaxies have intrinsically steeper radio spectra, because of increased IC losses from the CMB.  The rest-frame spectral slope $\alpha^{\prime}$ of normal galaxies at 1.4 GHz and asymptotes at $\sim 1.1$ by $z \approx 2 - 3$, when losses are dominated by IC for our injection spectrum of CRs with $p = 2.2$.  There is much less intrinsic rest-frame evolution of starburst radio spectra; even at $z \approx 5$, $\alpha^{\prime}$ increases only by 0.1 for the weakest compact starbursts.  The observable $\alpha^{1.4~\GHz}_{610~\MHz}$ shows much more pronounced evolution with redshift for starbursts, increasing by $0.1$ to $z \approx 1 - 2$, and $0.2$ to $z \approx 4 - 5
$.  This effect arises simply because we predict the CR electron and positron spectra steepen with rest-frame frequency as synchrotron and IC losses become stronger \citep[e.g.,][LTQ]{Thompson06}: at higher redshift and fixed observing frequency, we are seeing higher energy electrons and positrons.

An important effect that we do not consider is the thermal free-free radio emission.  This will set a minimum total observed radio emission that is directly proportional to the star-formation luminosity.  Thus, the true total radio deficit at GHz will not be as big as the synchrotron-only deficits plotted in Figures~\ref{fig:LFIRRadioRest} and \ref{fig:FRCEvolution}.  Thermal free-free emission will also flatten the spectrum, especially at high frequencies.  However, free-free emission is much fainter than the synchrotron luminosity at GHz frequencies, except in the faintest star-forming galaxies \citep{Condon92,Hughes06}.  

\subsection{The Radio Excess (or Deficit) of Puffy Starbursts: Submillimeter Galaxies}
\label{sec:Submillimeter}
\label{sec:Puffy}
As shown in Figure~\ref{fig:LFIRRadioRest}, puffy starbursts fall on a linear FIR-radio correlation of their own (dotted lines), in line with observations of SMGs \citep{Kovacs06,Murphy09}.  The radio luminosity of these galaxies is nonetheless the result of a conspiracy, between IC losses on starlight, which decrease the radio luminosity, and the enhanced radio emission from secondary electrons and positrons, and the $\nu_C$ effect (\S~\ref{sec:Theory}).  The variation in $L_{\rm TIR}^{\prime}/L_{\rm radio}^{\prime}$ for puffy starbursts alone is usually less than a factor of $2$ over the range $0.1~\gcm2 \le \Sigma_g \le 10~\gcm2$ (see Table~\ref{table:Models}; in most variants the variation is $\sim 1.6$).  Like compact starbursts, escape plays essentially no role in most of the models, except that winds can slightly decrease the radio emission in relatively tenuous $\Sigma_g = 0.1~\gcm2$ ($\Sigma_{\rm SFR} \approx 2 - 4~\Msun~\kpc^{-2}~\yr^{-1}$) starbursts (footnote~\ref{ftnt:WindVariant}).

\subsubsection{$B \propto \Sigma_g^{0.7-0.8}$: Radio-Bright Puffy Starbursts}
It is plain from Figure~\ref{fig:LFIRRadioRest} (\emph{left panel}) that the normalization of the puffy starburst FRC (dotted lines) is different than the FRC of the compact starbursts and normal galaxies (solid lines) when $B \propto \Sigma_g^{0.7}$.  We show in Table~\ref{table:Models} that models with $B \propto \Sigma_g^{0.7-0.8}$ have radio-bright puffy starbursts compared to the observed local FRC, by a factor of $2 - 4$ at $z = 0$.  Like the compact starbursts, puffy starbursts show little rest-frame evolution in $L_{\rm TIR}^{\prime}/L_{\rm radio}^{\prime}$, except at relatively low surface densities ($\Sigma_g = 0.1~\gcm2$; $\Sigma_{\rm SFR} \approx 2 - 4~\Msun~\kpc^{-2}~\yr^{-1}$), where they become radio dim at high redshifts because of IC losses on CMB photons.  Therefore, we predict that puffy starbursts, which are mainly observed at high $z$, have \emph{intrinsically} different radio properties not caused by their redshift.

We propose a natural explanation of this radio excess in the framework of LTQ (\S~\ref{sec:Theory}).  In dense starbursts, protons are efficiently converted into secondary electrons and positrons through inelastic proton-proton scattering, which contribute to the synchrotron emission.  Furthermore, the $\nu_C$ effect increases $L_{\rm TIR}^{\prime}/L_{\rm radio}^{\prime}$ for starbursts which have larger $B$.  Compact starbursts lying on the $z = 0$ FRC balance these effects with increased bremsstrahlung, ionization, and IC losses, which compete with the synchrotron losses and suppress the excess radio luminosity (\S~\ref{sec:Theory}).  In puffy starbursts with relatively low volume densities compared to their compact cousins at fixed $\Sigma_g$, however, bremsstrahlung and ionization are not strong enough to compensate for these effects.  Only synchrotron and IC losses remain, upsetting the conspiracy.  The radio excess predicted by this picture is systematically greater when using the K98 star-formation relation, because the IC loss rate on starlight is smaller at fixed surface density by a factor of 2.5 to 11 from weak starbursts to the densest starbursts.  This freedom to vary the radio-excess with $B$ does not exist in the standard calorimeter model, where all CR electron/positron energy goes into synchrotron emission, and the radio emission saturates.  The balance between synchrotron and the other forms of cooling can be changed in starbursts to alter the normalization of the FRC, even though escape is negligible.

There is a reason to expect our allowed $B \propto \Sigma_g^{0.7-0.8}$ specifically: radiation pressure may drive turbulence and enhance the magnetic field until its energy density is comparable to radiation \citep[e.g.,][]{Thompson08}.  If the K98 relation holds, then since the magnetic energy density scales as $U_B \propto U_{\rm ph}$, $B \propto \Sigma_g^{0.7}$, and if the B08 relation holds, then $B \propto \Sigma_g^{0.85}$.  This explanation is more problematic for the B07 relation and $B \propto \Sigma_g^{0.7}$. \footnote{However, we have scaled $B$ to the \emph{total} Milky Way magnetic field strength of $6~\muGauss$ near the Solar Circle.  If most of the magnetic field strength in starbursts is driven by turbulence, perhaps scaling to the disordered Galactic magnetic field strength \citep[$\sim 2 \muGauss$;][]{Beck01} near the Solar Circle would make more sense, because the ordered magnetic field arises from a different process.  The lower normalization for $B$ then would partly compensate for the steeper dependence on $\Sigma_g$.}  However, while we assume that there is a parametrization for $B$ that applied to both compact starbursts and puffy starbursts, the radio excess should arise more generally.  The radio excess arises simply because synchrotron cooling time is shorter than the bremsstrahlung and ionization cooling times in puffy starbursts, but longer in compact starbursts at fixed $\Sigma_g$: at fixed $\nu^{\prime}$, $t_{\rm brems} \propto n^{-1} \propto h / \Sigma_g$ and $t_{\rm ion} \propto B^{-1/2} n^{-1} \propto h / (B^{1/2} \Sigma_g)$.  We therefore expect there to be a radio excess with respect to the FRC in any puffy starburst with a strong enough magnetic field, with the exact enhancement depending on magnetic field strength because of the remaining competition, after ionization and bremsstrahlung are sub-dominant, from the IC losses on starlight.\footnote{If the magnetic field strength does not go very roughly as $B \propto \Sigma_g^{0.7}$, however, there will be more scatter in $L^{\prime}_{\rm TIR}/L^{\prime}_{\rm radio}$ for puffy starbursts at fixed $h$ and $z$.  SMGs would not form their own FRC if $B$ was the same for all SMGs regardless of $\Sigma_g$, or if $B$ increased \emph{very} steeply with $\Sigma_g$, because there still must be a conspiracy with IC losses off starlight.  Since SMGs do appear to form their own FRC \citep[e.g.,][]{Murphy09,Michalowski09}, this could be evidence specifically that their magnetic field strengths increase with $\Sigma_g$ (or $\rho$).}  

Our results imply that a moderate radio excess at the factor of $\sim 3$ level alone is not a safe indicator of the presence of a radio-loud AGN, especially at high redshifts where SMGs are observed.  While radio excess with respect to the local FRC has been suggested as a selection criterion for radio-loud AGNs \citep[e.g.,][]{Yun01,Yang07,Sajina08}, our models with $\Sigma_g^{0.7-0.8}$ imply that $q_{\rm FIR} \approx 1.7 - 2.0$ for puffy starbursts powered by star-formation alone.  A radio excess is inexplicable in our models only when the source is an order of magnitude brighter ($q_{\rm FIR} \la 1.5$) in the radio than predicted from the FIR emission.  While SMGs are relatively rare and may not be a problem in small samples, we recommend that other means be used to be sure that the radio-excess is caused by an AGN, such as a flat radio spectrum, radio morphology, mid-IR colors, or the presence of strong X-ray emission \citep[see also][]{Murphy09}.

\subsubsection{$B \propto \rho^{0.5 - 0.6}$: Radio-Dim Puffy Starbursts}
A magnetic field dependence of $B \propto \rho^{0.5}$ appears to hold for Galactic molecular clouds \citep[e.g.,][]{Crutcher99}, and LTQ found that the FIR-radio correlation was consistent with this magnetic field dependence.  The existence of galaxies with different scale heights allows us to distinguish the two possibilities for magnetic field scaling.  If $B \propto \Sigma_g^a$, then the magnetic field strength will be the same for all galaxies with the same $\Sigma_g$, regardless of scale height.  In models with $B \propto \rho^{0.5 - 0.6}$, by contrast, the magnetic field strength is weaker in puffy starbursts than in compact starbursts with the same $\Sigma_g$.  

As seen in Figure~\ref{fig:LFIRRadioRest} (\emph{right panel}, dotted lines), puffy starbursts again form their own FRC.  In models with $B \propto \rho^{0.5 - 0.6}$, they are radio dim compared to the $z \approx 0$ FRC.  We show in Table~\ref{table:Models} that the normalization of the FRC is radio-dim by a factor of $\sim 1.2 - 2.0$.  

We can explain this in the LTQ theory of the FRC as well.  The magnetic field strength must increase more slowly with $\rho$ than with $\Sigma_g$ to reproduce the $z \approx 0$ FRC, because $h \propto \Sigma_g / \rho$ is 10 times smaller in compact starbursts than normal galaxies and puffy starbursts.  Compact starbursts are highly compressed with respect to normal galaxies, so they have strong magnetic fields and synchrotron radio emission is strong enough to compete with the other losses.  Puffy starbursts are not compressed, so that their magnetic fields are weak and synchrotron losses cannot keep up with IC losses, nor with bremsstrahlung and ionization as 1.4 GHz emission traces ever lower electron energies at higher magnetic field strengths.  Puffy starbursts therefore turn out to be radio dim compared to compact starbursts on the $z \approx 0$ FRC, if $B \propto \rho^{0.5 - 0.6}$.  As before, the B07 star-formation relation predicts greater IC losses and therefore weaker radio emission.  

Because of the claims in the literature that SMGs are radio bright, we do not favor these models.  The suggested relative radio brightness of high-$z$ SMGs therefore provides some evidence that, in fact, $B \propto \Sigma_g^{0.7 - 0.8}$ rather than $B \propto \rho^{0.5 - 0.6}$.  However, the matter of whether high-$z$ SMGs are in fact radio bright is not yet settled.  Although LTQ concluded that $B$ must increase dramatically from normal galaxies to dense starbursts (\S~\ref{sec:Theory}), they were unable to distinguish between these two possibilities with the $z \approx 0$ FRC alone.  For this reason, high-$z$ starbursts and their qualitatively different morphologies compared to those at $z \approx 0$ can distinguish theories of the FRC.  

\subsubsection{Spectral slopes}
A prediction of all of our variants is that puffy starbursts like submillimeter galaxies should have steep non-thermal radio spectra, with $\alpha \approx 0.8 - 1.0$ (see Table~\ref{table:Models}).  The steep spectra are caused by strong synchrotron cooling in the $B \propto \Sigma_g^{0.7-0.8}$ case and the relatively stronger IC cooling off starlight in the $B \propto \rho^{0.5 - 0.6}$ case.  In general, puffy starbursts should have roughly the same $\alpha$ as normal galaxies in the local universe, which tends to be somewhat higher ($\alpha \approx 0.7 - 1.0$) than in compact starbursts ($\alpha \la 0.7$).  The slope should hold even out to extremely high $\Sigma_g$, as long as starbursts are puffy.  In contrast, we find that $\alpha \approx 0.5$ in compact starbursts, because of efficient ionization and bremsstrahlung losses, which flatten the equilibrium CR spectrum because of their energy dependence.  As we note in LTQ, our predicted spectral index for normal galaxies is somewhat too high, and this difference in $\alpha$ may carry over to the puffy starbursts.  However, the significant difference in $\alpha$ between compact and puffy starbursts should remain as a general prediction of our model: compact starbursts should have flatter spectra than puffy starbursts.

The high spectral slopes can be observed either with direct measurements of multifrequency data of individual submillimeter galaxies, or with single frequency observations at a variety of redshifts.  There are relatively few measurements of $\alpha$ for submillimeter galaxies specifically; faint radio sources have $\alpha \approx 0.5 - 0.7$ \citep{Huynh07,Bondi07}, though that sample includes both compact starbursts and AGNs.  \citet{Sajina08} do find that $\alpha_{610~\MHz}^{1.4~\GHz} \approx 0.8$ for SMGs, comparable to our predictions.  They also find that submillimeter galaxies have a radio-excess, in agreement with Figure~\ref{fig:LFIRRadioRest}.  More recently, \citet{Ibar10} found an average $\alpha_{610~\MHz}^{1.4~\GHz} \approx 0.75 \pm 0.06$, which is somewhat flatter than our models.  These spectral slopes are not different from normal star-forming galaxies, but are noticeably steeper than local ULIRGs \citep{Clemens08}.  However, we do not account for free-free absorption, which probably flattens the spectra of local ULIRGs like Arp 220 at low frequency \citep{Condon91}, and is not well understood in SMGs.

Since puffy starbursts have steeper spectra than compact starbursts, we expect their inferred $L_{\rm TIR}^{\prime}/L_{\rm radio}^{\prime}$ will increase with redshift: if the true radio spectral slopes of SMGs are greater than the assumed $\alpha$ by $\Delta\alpha$, they will appear to become radio dimmer by a factor $(1 + z)^{\Delta \alpha}$, or up to $\sim 40\%$ at $z = 2$.  

\begin{figure}
\centerline{\includegraphics[width=9cm]{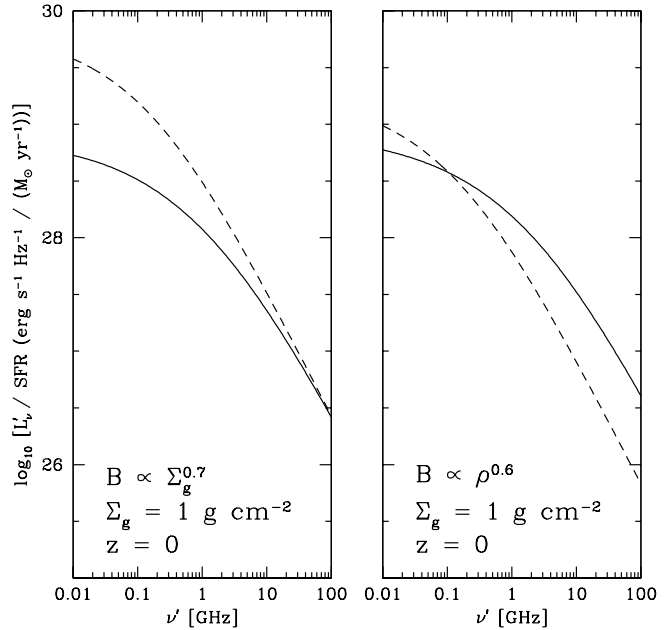}}
\figcaption[figure]{The rest-frame synchrotron radio spectra of starbursts with $\Sigma_g = 1~\gcm2$, using the B07 star-formation law. Puffy starbursts are dashed, while compact starbursts are solid. These spectra do not include thermal absorption (at low frequencies, $\la 1\ \GHz$) or emission (at high frequencies, $\ga 30\ \GHz$).  \label{fig:StarburstSpectra}}
\end{figure}

In Figure~\ref{fig:StarburstSpectra}, we show the expected radio synchrotron spectra of starburst galaxies, without correcting for thermal absorption or thermal emission.  At a rest-frame frequency of 1 GHz, puffy starbursts (dashed) have steeper radio spectra than compact starbursts (solid).  Note that at high frequencies ($\nu^{\prime} \ga 10\ \GHz$), the ratio of the radio luminosities per unit star formation of the compact and puffy starbursts asymptotes to a value set by the ratio of $U_B$ and $U_{\rm ph}$ in these starbursts.  At these high frequencies, only synchrotron and IC cooling are effective, and IC cooling would be the same for puffy and compact starbursts because of the Schmidt Law (\S~\ref{sec:Theory}).  For $B \propto \Sigma_g^a$, $U_B$ is the same for puffy and compact starbursts, but for $B \propto \rho^a$, puffy starbursts have much smaller $U_B$.  Thus, measurements of the synchrotron radio emission of SMGs at high $\nu^{\prime}$ could determine the magnetic field strength of SMGs and determine which scenario applies. 

Of course, there are unlikely to be two perfectly distinct populations of compact starbursts and puffy starbursts.  Instead, there may be a continuum variation in scale heights from tens to thousands of parsecs.  We would then expect to see a larger scatter, both in $L_{\rm TIR}^{\prime}/L_{\rm radio}^{\prime}$ and $\alpha$, in a full sample of both the most compact and the most puffy starbursts.  \citet{Murphy09} find that submillimeter galaxies do have a larger scatter in $q_{\rm TIR}^{\prime}$ than other galaxies.  However, \citet{Ibar10} find a relatively small scatter of $\sim 0.3$ in SMG radio spectral index.  Importantly, for larger $h$, both $L_{\rm TIR}^{\prime}/L_{\rm radio}^{\prime}$ and $\alpha^{\prime}$ asymptote as CR electron and positron losses are entirely determined by synchrotron and IC; $h \approx 1~\kpc$ starbursts are already near this limit.  In other words, for arbitrarily large $h$, the radio excess with respect to the $z \approx 0$ FRC asymptotes to a value of $\sim 5 - 10$, depending on the assumed Schmidt Law, and the radio spectral slope asymptotes to $\sim 1.1$ for $p = 2.2$.

\subsection{Synchrotron Radio Emission as a Star-Formation Tracer}
\label{sec:SFTracer}

\begin{figure}
\centerline{\includegraphics[width=9cm]{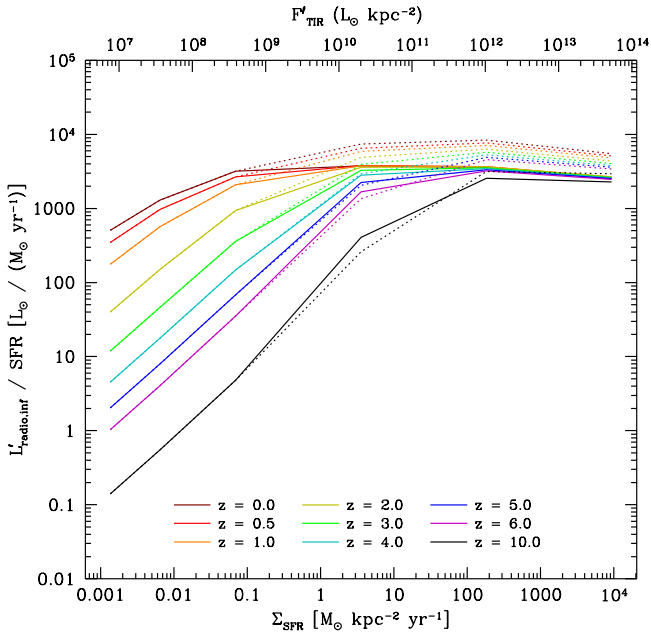}}
\figcaption[figure]{The relationship between the inferred rest-frame radio emission and star-formation, using the B07 star formation law, $B \propto \Sigma_{\rm SFR}^{0.7}$, and no winds.  Solid lines have $h = 100~\pc$ for starbursts, while dotted lines have $h = 1~\kpc$ for starbursts.  A flat line would indicate that radio emission is directly proportional to $\Sigma_{\rm SFR}$.  We see that at low $\Sigma_g$, the radio flux underestimates the star-formation rate even at $z = 0$, because of CR electron escape (the low-$\Sigma_g$ conspiracy of \S~\ref{sec:Theory}).  The radio flux overestimates the star-formation rate for puffy starbursts, because the high-$\Sigma_g$ conspiracy is unbalanced (\S~\ref{sec:Theory}; \S~\ref{sec:Puffy}).  Finally, the radio emission is suppressed at high redshift, partly because of IC losses off the CMB, and partly because $\alpha \ga 0.7$ for puffy starbursts and normal galaxies ($\alpha = 0.7$ assumed here for the k-correction).\label{fig:SFRRadio}}
\end{figure}

Relying on the FRC, a number of studies have used the GHz radio emission as a tracer of star-formation \citep[e.g.,][]{Cram98,Mobasher99,Haarsma00,Carilli08,Seymour08,Garn09}.  Radio emission has the advantage that it is unaffected by dust obscuration, making it potentially very useful in starbursts, and therefore for most of the star-formation at $z \ga 1$ \citep[e.g.,][]{Chary01,LeFloch05,Dole06,Magnelli09,Pascale09}.  Our models let us evaluate the theoretical basis for radio as a SFR indicator at all redshifts.  We show the predicted radio emissivity as a function of star formation in Figure~\ref{fig:SFRRadio}.  

At low surface densities ($\Sigma_g \la 0.01~\gcm2$; $\Sigma_{\rm SFR} \la 0.06~\Msun~\kpc^{-2}~\yr^{-1}$), synchrotron radio has a non-linear dependence on star-formation at all redshifts.  The weak radio emission is caused by electrons escaping their host galaxies, as inferred by \citet{Bell03} and discussed in LTQ.  That is, normal galaxies are not perfect electron calorimeters, so the radio emission is not a reliable star-formation tracer at low $\Sigma_{\rm SFR}$.  At higher redshift, synchrotron emission is diminished by Inverse Compton off the CMB.  For a Milky Way-like galaxy, we find that radio is a good star-formation tracer at $z \la 1$, but underestimates it significantly by $z \ga 2$ (eq. \ref{eqn:B07zCrit}).  Already galaxies with star formation rates similar to Galactic levels are beginning to be observed in the radio at high redshift \citep{Garn09}, so that the IC suppression may soon be observed.  However, IC losses off the CMB should not be important even in the weakest starbursts until $z \ga 4$, as is also visible by the redshift evolution of the FRC in Figure~\ref{fig:FRCEvolution}.

Radio emission does grow linearly with star-formation rate between normal galaxies with $\Sigma_g = 0.01~\gcm2$ (at $z \la 2$) and compact starbursts.  Therefore, it serves as an acceptable star-formation indicator for these galaxies.  However, if $B \propto \Sigma_g^{0.7-0.8}$, puffy starbursts like SMGs have about $2 - 4$ times the radio emission at any given star-formation rate than compact starbursts.  Therefore, we expect that if $B \propto \Sigma_g^{0.7-0.8}$, the usual radio emission estimate based on the $z \approx 0$ FRC will \emph{overestimate} their star formation rates by a factor of $\sim 2 - 4$.  The excess is greatest at $\Sigma_g \approx 1~\gcm2$ corresponding to $\Sigma_{\rm SFR} \approx 40 - 200~\Msun \kpc^{-2} \yr^{-1}$, typical of observed SMGs.  Among the puffy starbursts themselves, the radio emission grows linearly with star-formation rate.  If instead $B \propto \rho^{0.5 - 0.6}$, the radio emission will underestimate the star-formation rate.  However, because the SMGs lie on their own FRC, there is little real redshift evolution in the radio emissivity of puffy starbursts, because of the buffering provided by IC losses off starlight (see \S~\ref{sec:FRCEvolution}, Appendix~\ref{sec:CMBRadioDim}).  

Assuming a spectral slope $\alpha = 0.7$ will also underestimate the radio emissivity, since these galaxies can have steep radio spectra.  This explains the apparent evolution with $z$ of puffy starbursts in Figure~\ref{fig:SFRRadio}: we have applied a k-correction using a typically employed $\alpha = 0.7$ when, in fact, the true synchrotron spectra are steeper.  If the radio star-formation tracer could be calibrated to the special conditions in puffy starbursts like SMGs, taking into account their different scale height, radio emissivity, and spectral slopes, we predict that radio would be a more accurate star-formation tracer for them.

\begin{figure}
\centerline{\includegraphics[width=9cm]{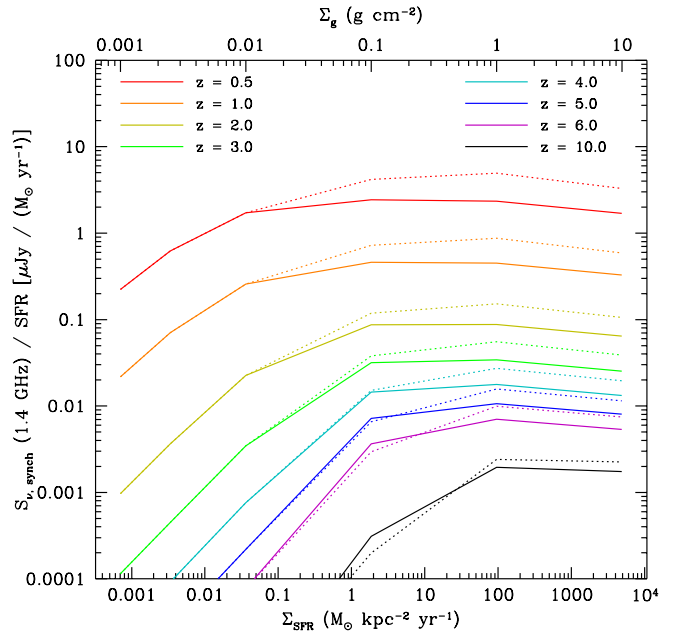}}
\figcaption[figure]{The \emph{observed} 1.4 GHz synchrotron radio flux per unit star-formation as a function of redshift, using the B07 star formation law, $B \propto \Sigma_{\rm SFR}^{0.7}$, and no winds.  EVLA will have a sensitivity of $\sim 1\ \muJy$ and SKA will have a sensitivity of $\sim 20\ \nJy$.\label{fig:GHzRadioFluxes}}
\end{figure}

We can also directly calculate the \emph{observed} GHz radio flux density $S_{\nu}$ from synchrotron emission of star-forming galaxies.  We show the predicted flux density per unit star-formation at observer-frame 1.4 GHz in Figure~\ref{fig:GHzRadioFluxes}.  Surveys with the Expanded Very Large Array (EVLA) will have a continuum sensitivity of $\sim 1\ \muJy$ at frequencies of 1 - 50 GHz, and it should be able to directly detect galaxies with $\Sigma_g \ga 0.01\ \gcm2$ and star-formation rates of $1\ \Msun\ \yr^{-1}$ out past $z \approx 0.5$.  As stated in \citet{Murphy09c}, a starburst like M82 with SFR $\approx 3\ \Msun\ \yr^{-1}$ will become undetectable past $z \approx 1$.  However, the buffering effect we emphasize in this paper preserves the radio emission of dense starbursts at high $z$, so that bright starbursts will be detectable further: starbursts with SFR $\ga 100\ \Msun\ \yr^{-1}$ will be detectable with EVLA at 1.4 GHz out to $z \approx 4 - 5$ and the most intense starbursts (SFR $\ga 1000\ \Msun\ \yr^{-1}$ and $\Sigma_g \ga 1~\gcm2$) will be detectable out past $z \approx 10$ in synchrotron emission.  \citet{Murphy09c} predicts the Square Kilometer Array (SKA) will be sensitive to star-forming galaxies with flux densities of $\sim 20\ \nJy$.  If this sensitivity is attained, then Milky Way-like galaxies ($\Sigma_g \approx 0.01\ \gcm2$; SFR $\approx 1\ \Msun\ \yr^{-1}$) will be directly detectable at 1.4 GHz in synchrotron emission out to $z \approx 2$, and starbursts with SFR $\approx 1\ \Msun\ \yr^{-1}$ will be detectable at 1 GHz beyond $z \approx 3$.  Even at $z \approx 10$, the synchrotron emission of dense, compact starbursts ($\Sigma_g \ga 1\ \gcm2$) with SFR greater than $10\ \Msun\ \yr^{-1}$ should be detectable at 1.4 GHz with SKA (models with $B \propto \rho^{0.5-0.6}$ have radio-dim puffy starbursts, and these are detectable at $z = 10$ with the SKA only for SFR greater than $20 - 60\ \Msun\ \yr^{-1}$).  By contrast, \citet{Murphy09c} found a sensitivity of $25\ \Msun\ \yr^{-1}$, based on the free-free emission; this limit will apply to normal galaxies and weak starbursts where the synchrotron emission is suppressed.  These sensitivities assume that natural confusion, in which radio sources overlap, will not hamper the SKA; estimates for the natural confusion limit vary from nJy to $\muJy$ levels \citep[e.g.,][]{Jackson04,Condon09,Murphy09c}.

SKA and EVLA will also have good spectral coverage, which may help measurements of the spectral index.  If a galaxy is detected at $5 \sigma$ ($\sim \muJy$ for EVLA and $20\ \nJy$ for SKA; \citealt{Murphy09c}) at two different frequencies $\nu_1$ and $\nu_2$ with $\nu_2 / \nu_1 = 5$, then $\alpha_1^2$ can be constrained to $\sim 0.1 - 0.15$ at the $1 \sigma$ level.  EVLA will be better at high observer-frame frequencies (1 - 50 GHz).  A problem for the EVLA will be the increasing fraction of thermal emission, which is expected to dominate the emission of starbursts at $\nu^{\prime} \approx 30\ \GHz$.  EVLA will therefore not easily measure the nonthermal spectral indices of galaxies at high z.  On the other hand, the SKA will face free-free absorption when observing low-z starbursts; for example, Arp 220 may be optically thick even at 1 GHz \citep{Condon91}.  At high redshift, however, the rest-frame frequencies SKA will observe will be less affected by free-free absorption.

\subsection{SMGs, The Radio Background, and Radio Source Counts}
\label{sec:RadioBackground}
Star-formation in galaxies over cosmic time produces a diffuse radio synchrotron background.  Our models indicate that submillimeter galaxies and other puffy starbursts ought to have enhanced radio emission.  In principle, this means that the synchrotron radio background could be up to $\sim 2 - 4$ times higher than usually predicted from the Cosmic Infrared Background (CIB) and a naive application of the $z \approx 0$ FRC.  The magnitude of this implied radio excess is interesting, because ARCADE2 recently reported an excess radio background at 3~\GHz, about five times higher than expected from star formation \citep{Fixsen09,Seiffert09}.  \citet{Singal09} found that constraints on Inverse Compton emission require the excess to come from regions with galactic-level ($\ga \muGauss$) magnetic fields, and suggest an evolution of the FRC as the source of the reported excess.  However, while SMGs are individually very bright and contribute much to the cosmic star-formation rate at $z \ga 2.5$, they are not typical starbursts \citep{Bavouzet08}.  Instead, they seem to represent a transient phase that can survive about 100~\Myr~before their gas is depleted \citep{Tacconi06,Pope08}.

We can estimate the total radio background enhancement by scaling to the contribution of SMGs to the CIB, which is largely reprocessed  starlight from galaxies at $z \ga 1$ \citep[e.g.,][]{Dole06,Devlin09}, and adjusting by the SMG radio excess.  Submillimeter galaxies do provide the majority of light at $\sim 850 \mum$, but this is only a small fraction of the total IR background.  At the peak of the CIB ($\sim 160 \mum$), submillimeter galaxies provide $\la 10\%$ of the total power, and possibly only $\sim 2\%$ \citep{Chapman05,Dye07}.  Optimistically, the radio excess from SMGs would be $10\% \times (4 - 1) \approx 30\%$.  This is significant, but not enough to explain the very large ARCADE2 excess.  More conservatively, the excess is more likely $5\% \times 2 \approx 10\%$, and could be as little as a few percent.  

Since the number of radio sources down to several $\muJy$ is well known, we can estimate the fraction of the expected radio background comes from bright SMGs.  \citet{Dole06} find a TIR background of about $24~\nWBackUnits$; from the normalization of the FIR-radio correlation we expect a 1.4 GHz background of $\nu I_{\nu} \approx 2.6 \times 10^{-5}~\nWBackUnits$.  \citet{Chapman05} found an average radio flux density of $S_{1.4} \approx 75\ \muJy$ for bright SMGs ($S_{\rm 850\ \mu m} \ga 5\ \mJy$).  The number counts of $75\ \muJy$ sources imply that they have a density of $\sim 1500\ \deg^{-2}$, contributing roughly $\sim 5.1 \times 10^{-6}~\nWBackUnits$ to the 1.4 GHz radio background \citep[e.g.,][]{Gervasi08}.  Bright SMGs have an approximate density of $\sim 600\ \deg^{-2}$ \citep[e.g.,][]{Wang04,Coppin06}, so they constitute about $\sim 40\%$ of the background from $75\ \muJy$ sources, or $\sim 8\%$ of the expected 1.4 GHz background from star-formation.  This is roughly in line with our estimate of $\sim 10\%$, although the uncertainties are large enough that it could be consistent with SMGs lying on the FRC.  In any case, it is fairly clear that bright SMGs are not the source of the ARCADE excess.

The total excess could be greater if most starbursts at $z \ga 1$ are puffy, and not just SMGs.  We do not expect this simply because most current studies show that the local FRC \emph{does} hold out to high redshift for most observed galaxies and starbursts \citep[e.g.,][]{Murphy09,Younger09,Garn09}.  The spectral slope would also be a problem: ARCADE2 inferred $\alpha = 0.6$, while we predict a spectral slope $\alpha \ga 0.7$, steeper than local compact starbursts.  Resolved radio sources in the range $50\ \muJy \la S_{1.4} \la 1\ \mJy$, which are expected to be star-forming galaxies \citep{Danese87,Condon89,Benn93}, contribute about $1.1 \times 10^{-5}\ \nWBackUnits$ to the 1.4 GHz radio background \citep{Gervasi08}, which is consistent with the total energetics expected from the FRC.  \citet{deZotti10} also find consistency between the FRC and the observed radio source counts at $\sim 30\ \muJy$.  Stacking studies of fainter radio sources have given somewhat ambiguous results on whether the FRC applies \citep{Boyle07,Beswick08}, but \citet{Garn09b} find no large evolution in the FRC down to $S_{1.4} \approx 20\ \muJy$.  Any large radio excess would have to come either from a new population of low luminosity galaxies or very high-$z$ galaxies (see Figure~\ref{fig:GHzRadioFluxes}); extrapolations of the higher flux source populations do not predict a large radio excess \citep[e.g.][]{Gervasi08}.  

Nonetheless, our work indicates that a radio excess from SMGs can be significant.  Conversely, if the luminosity function of galaxies was very steep, most galaxies could be intrinsically radio-dim with respect to the FRC, because of IC losses off the CMB (see \S~\ref{sec:FRCEvolution} and Figures~\ref{fig:LFIRRadioRest} and~\ref{fig:FRCEvolution}).  Then the FRC would overestimate the strength of the radio background.  However, most star-formation at high $z$ is believed to have occurred in starbursts \citep{LeFloch05,Dole06}, so this possibility is unlikely.  

\section{Summary and Cavaets}
\label{sec:Summary}
We have applied the theory of LTQ to predict the FRC for redshifts $0 \le z \le 10$.  We use one-zone models of galaxies and starbursts with CR injection, cooling, and escape to predict the equilibrium, steady-state radio spectra of galaxies and starbursts over the entire range of the FRC.  Our goals were to determine how and why the low- and high-$\Sigma_g$ conspiracies crucial to the $z \approx 0$ FRC (\S~\ref{sec:Theory}) affect the FRC at high redshift, and to provide a quantitative model for predicting the critical redshifts at which galaxies deviate from the $z \approx 0$ FRC.  We find the following:

\begin{enumerate}
\item For compact starbursts ($h \approx 100~\pc$), we find relatively little evolution in the FIR-radio correlation out to $z \approx 5 - 10$ (Figure~\ref{fig:LFIRRadioRest}).  This is partly because the magnetic energy density in galaxies is strong enough to dominate the CMB even at high redshifts.  However, the high-$\Sigma_g$ conspiracy (\S~\ref{sec:Theory}) also acts as a buffer against IC losses off the CMB; the increased IC losses must compete with the already present bremsstrahlung, ionization, and IC off starlight in addition to synchrotron losses.  The rest-frame radio spectral slope $\alpha^{\prime}$ at fixed $\nu^{\prime}$ does not change with $z$, but the observed $\alpha$ at fixed $\nu$ increases because the non-thermal synchrotron radio spectrum steepens at higher rest-frame frequency.

\item We derive in Appendix~\ref{sec:CMBRadioDim} the critical redshifts when Inverse Compton losses off the CMB suppress the radio luminosity of galaxies compared to the $z \approx 0$ FRC.  These relations are given for our standard model in equation \ref{eqn:B07zCrit}.  The non-thermal radio luminosity is suppressed severely in Milky Way-like galaxies ($\Sigma_{\rm SFR} \approx 0.06~\Msun~\kpc^{-2}~\yr^{-1}$) at $z \approx 2$ and the weakest compact starbursts at $z \ga 5$.  The spectrum at GHz steepens to $\alpha \approx 1$ because of these enhanced IC losses.  Nonetheless, the low-$\Sigma_g$ conspiracy (\S~\ref{sec:Theory}) also acts to prevent the radio emission from steeply falling with redshift, since Inverse Compton losses off the CMB must be more efficient than diffusive escape, not just synchrotron losses (see \S~\ref{sec:FRCEvolution}). 

\item LTQ found that the $z = 0$ FRC demands that $B$ scales with $\rho$ or $\Sigma_g$ in galaxies lying on the Schmidt law.  In models with $B \propto \Sigma_g^{0.7-0.8}$, we find that puffy starbursts with $h = 1~\kpc$ such as SMGs are radio bright compared to the $z = 0$ FRC by a factor of $\sim 2 - 4$.  This follows from a breakdown of the high-$\Sigma_g$ conspiracy (\S~\ref{sec:Theory}): bremsstrahlung and ionization cooling are weak in puffy starbursts relative to the compact starbursts that predominate in the $z = 0$ universe.  In contrast, in models with $B \propto \rho^{0.5 - 0.6}$, we find that puffy starbursts are radio dim compared to the observed FRC, because of weak synchrotron cooling relative to the IC losses.  Since several studies have reported radio excesses for SMGs, we favor the $B \propto \Sigma_g^{0.7-0.8}$ scaling; however the issue of whether SMGs are radio-bright is still not fully resolved.  In either case, puffy starbursts show little true evolution with $z$, though they may appear to have fainter rest-frame radio luminosities at high $z$ because of their steep spectra.  Puffy starbursts inevitably have high $\alpha$ ($\ga 0.7$), since bremsstrahlung and ionization losses are weak with respect to synchrotron and IC.  A key prediction of our scenario with $B \propto \Sigma_g^{0.7-0.8}$ is that the variations in $L_{\rm TIR}^{\prime}/L_{\rm radio}^{\prime}$ will be correlated with scale height at fixed $\Sigma_{\rm SFR}$, since the radio-excess in our models is a direct consequence of the large CR scale height and the small bremsstrahlung and ionization losses it causes.  Radio-excess (low $q$) puffy starbursts will have steeper radio spectral slopes (bigger $\alpha$), larger velocity dispersions $\sigma$ compared to their rotation speeds $v_{\rm circ}$, and possibly moderately cooler dust temperatures (smaller $T_{\rm dust}$).

\item As previously expected, radio emission can be a poor tracer of star formation in low surface density galaxies, because of electron escape and IC losses off the CMB.  For our preferred $B \propto \Sigma_g^{0.7-0.8}$ scaling, radio emission overestimates star-formation rate by a factor of $2 - 4$ in puffy starbursts.  Star-formation rate is underestimated by synchrotron radio emission with the $B \propto \rho^{0.5 - 0.6}$ scaling.

\item While SMGs may be individually radio bright compared to the local FRC, they contribute a relatively small fraction of the Cosmic Infrared Background and the total star-formation luminosity of the Universe.  This means that they enhance the star-formation radio background by $\la 50\%$, and possibly around $\sim 10\%$, with respect to a naive application of the $z = 0$ FRC.
\end{enumerate}

As in LTQ, we did not exactly match the observed radio spectral slopes of galaxies, with $\alpha \approx 0.9 - 1.0$ in normal galaxies and $\alpha \approx 0.4 - 0.6$ in compact starbursts.  This will have a slight effect on the k-correction.  An error of $0.25$ in $\alpha$ should only affect inferred radio luminosity by 30\% at $z = 2$ and 60\% at $z = 5$.  Nevertheless, the prediction of steeper radio spectra in puffy starbursts with respect to compact starbursts at all relevant $z$ should be robust.

Our explanation for the small $L_{\rm TIR}^{\prime}/L_{\rm radio}^{\prime}$ ratio in submillimeter galaxies as a breakdown of the high-$\Sigma_g$ conspiracy is based purely on the steady-state spatially-averaged synchrotron emission, but the details of the FIR emission may also matter.  Throughout this paper, we have simply assumed that the bolometric FIR luminosity could be correctly inferred from observations, and have assumed the same UV opacity for all galaxies at all redshifts.  The total FIR emission is also likely to depend on the metallicity, and may be lower at high $z$ for the lowest surface density galaxies.  The exact far-infrared SED is important in determining the FIR emission when observations have only been made at only a few wavelengths.  The presence of AGNs, a different IMF at high $z$, and selection biases may also affect the inferred $q$ of SMGs.

We did not include the effects of galaxy evolution on the CR spectrum in our models.  We argued in LTQ that it should not matter for quiescent spirals or for extreme starbursts, because the CR lifetime is much shorter than the time dependence of stellar populations.  However, galaxy evolution may play a role in weaker starbursts \citep{Lisenfeld96b} and in post-starburst galaxies \citep{Bressan02}.  Studies of merging normal galaxies and galaxies in clusters have indeed found that they are radio bright with respect to the FRC, possibly because of compression of magnetic fields or shock acceleration \citep{Gavazzi91,Miller01,Murphy09b}.  

We also assumed that the magnetic field strength at a given density does not depend on redshift.  It is not entirely clear how long normal galaxies take to build up their magnetic fields, or even what process is at work \citep[see the reviews in ][]{Widrow02,Kulsrud08}, though there are theoretical mechanisms that can rapidly generate strong magnetic fields.  Studies at $z \approx 2$ indicate that normal Milky Way-like galaxies had magnetic fields with similar strengths to the present \citep{Kronberg08,Bernet08,Wolfe08}.  At the very highest redshifts, magnetic field strengths might be weaker, because the seed fields were essentially zero compared to the present strengths.  Starbursts also may build their magnetic fields up in much shorter times than normal galaxies, and through a different process than normal galaxies \citep{Thompson09}.

Finally, we have used one-zone models, which are appropriate if the CRs sample all of the gas phases in each galaxy's ISM.  However, the ISM is known to be clumpy in the Milky Way, in compact starbursts like Arp 220 \citep[e.g.,][]{Greve09}, and even in the SMGs themselves \citep{Tacconi06}.  A full understanding of the FRC will probably require models that take into account the inhomogeneity of star-forming galaxies.

\acknowledgments
We thank Eric Murphy and Michal Micha{\l}owski for critical readings of the text, Eliot Quataert for many stimulating conversations, and the GALPROP team for making their code and its subroutines freely available.  GALPROP is available at http://galprop.stanford.edu.   We would like to thank Igor Moskalenko for sharing his group's estimates of the Milky Way's total $\gamma$-ray luminosity for normalization of our models of the FRC.  T.~A.~T. is supported in part by an Alfred P. Sloan Fellowship.

\appendix
\section{Derivation of Radio Suppression from CMB}
\label{sec:CMBRadioDim}

At high redshift, the nonthermal synchrotron radio luminosity of galaxies is suppressed by the CMB.  This is because the Inverse Compton process shortens the lifetime that CRs have to radiate synchrotron; equivalently, their energy goes into Inverse Compton photons instead of synchrotron radio.  Because of the shorter times the CR electrons and positrons have to radiate synchrotron, the redshift $z_{\rm crit}$ when the radio luminosity of a galaxy is quenched by a factor ${\cal Q}$ is given by 
\begin{equation}
\label{eqn:RadioSuppress}
{\cal Q} \frac{t_{\rm synch}}{t_{\rm loss}} (z = 0) = \frac{t_{\rm synch}}{t_{\rm loss}} (z = z_{\rm crit}).
\end{equation}
The loss time includes all cooling and escape losses, so that
\begin{equation}
\frac{t_{\rm synch}}{t_{\rm loss}} = 1.0 + \frac{t_{\rm synch}}{t_{\rm IC,\star}} + \frac{t_{\rm synch}}{t_{\rm IC,CMB}} + \frac{t_{\rm synch}}{t_{\rm brems}} + \frac{t_{\rm synch}}{t_{\rm ion}} + \frac{t_{\rm synch}}{t_{\rm diff}} + \frac{t_{\rm synch}}{t_{\rm adv}}
\end{equation}
In some regimes, some losses may be neglected.  For example, diffusive losses are unimportant in starbursts, as are IC losses off the CMB at $z = 0$.  Advective losses may be unimportant in normal galaxies and also for the densest starbursts.

We can immediately see that the presence of losses besides synchrotron and IC off the CMB prevent the suppression of radio.  In the case when no losses other than synchrotron and IC off the CMB exist, equation~\ref{eqn:RadioSuppress} reduces to
\begin{equation}
({\cal Q} - 1) \frac{U_B}{U_{\rm CMB} (z = 0)} + {\cal Q} = (1 + z_{\rm crit})^4.
\end{equation}
The presence of other loss processes, whether escape or cooling, requires a higher $z_{\rm crit}$ for a fixed quench factor ${\cal Q}$:
\begin{equation}
({\cal Q} - 1) \frac{U_B}{U_{\rm CMB} (z = 0)} \left(1.0 + \frac{t_{\rm synch}}{t_{\rm IC,\star}} + \frac{t_{\rm synch}}{t_{\rm brems}} + \frac{t_{\rm synch}}{t_{\rm ion}} + \frac{t_{\rm synch}}{t_{\rm diff}} + \frac{t_{\rm synch}}{t_{\rm adv}}\right)  + {\cal Q} = (1 + z_{\rm crit})^4.
\end{equation}
Essentially the IC losses off the CMB must compete with not only synchrotron losses, but every other cooling and escape process as well.  With the high-$\Sigma_g$ conspiracy, $t_{\rm synch} / t_{\rm loss}$ can be of order $10 - 20$, so the energy density of the CMB must be $\sim 10 - 20$ times greater to suppress the radio emission than would be naively expected.  

We give the loss times for each process in LTQ.  From those lifetimes, we can find the ratios of the synchrotron cooling timescale to the other loss timescales:
\begin{eqnarray}
\frac{t_{\rm synch}}{t_{\rm IC, CMB}} & = & 0.11 B_{10}^{-2} (1 + z)^4\\
\frac{t_{\rm synch}}{t_{\rm IC, \star}} & = & 19 B_{10}^{-2} \left(\frac{\Sigma_{\rm SFR}}{\Msun~\kpc^{-2}~\yr^{-1}}\right)\\
\frac{t_{\rm synch}}{t_{\rm brems}} & = & 98 f \nu_{1.4}^{\prime -1/2} B_{10}^{-3/2} h_{\rm kpc}^{-1} \left(\frac{\Sigma_g}{\gcm2}\right)\\
\frac{t_{\rm synch}}{t_{\rm ion}} & = & 15 f \nu_{1.4}^{\prime -1} B_{10}^{-1} h_{\rm kpc}^{-1} \left(\frac{\Sigma_g}{\gcm2}\right)\\
\frac{t_{\rm synch}}{t_{\rm diff}} & = & 1.5 \nu_{1.4}^{\prime -1/4} B_{10}^{-7/4}\\
\frac{t_{\rm synch}}{t_{\rm adv}} & = & 12 \nu_{1.4}^{\prime -1/2} B_{10}^{-3/2} h_{\rm kpc}^{-1} v_{300},
\end{eqnarray}
where $B_{10} = B / (10 \muGauss)$, $h_{\rm kpc} = h / (1 \kpc)$, $v_{300}$ is the wind speed in units of 300 $\kms$, and $\nu_{1.4}^{\prime}$ is the rest-frame frequency divided by 1.4 GHz.  

Our models parameterize $B$ in terms of the gas surface density $\Sigma_g$:
\begin{equation}
B_{10} =  \left\{ \begin{array}{ll}
			40 (\Sigma_g / \gcm2)^{0.7} & (B \propto \Sigma_g^{0.7})\\
			72 (\Sigma_g / \gcm2)^{0.8} & (B \propto \Sigma_g^{0.8})\\
			12 (\Sigma_g / \gcm2)^{0.5} h_{\rm kpc}^{-0.5} & (B \propto \rho^{0.5})\\
			22 (\Sigma_g / \gcm2)^{0.6} h_{\rm kpc}^{-0.6} & (B \propto \rho^{0.6}).
			\end{array} \right.
\end{equation}
Finally, the Schmidt Law allows us to convert between gas surface density $\Sigma_g$ and surface density of star-formation rate $\Sigma_{\rm SFR}$:
\begin{equation}
\left(\frac{\Sigma_g}{\gcm2}\right) =  \left\{ \begin{array}{ll}
\displaystyle
			0.078 \left(\frac{\Sigma_{\rm SFR}}{\Msun~\kpc^{-2}~\yr^{-1}}\right)^{0.71} & ({\rm K98})\\
\displaystyle
			0.048 \left(\frac{\Sigma_{\rm SFR}}{\Msun~\kpc^{-2}~\yr^{-1}}\right)^{0.59} & ({\rm B07}).
			\end{array} \right.
\end{equation}

We can approximately solve for the redshift $z_{\rm crit}$ when ${\cal Q} = 3$ for galaxies and starbursts in each of the scenarios in Table~\ref{table:Models}.  For the scenario with the B07 star-formation law, $B \propto \Sigma_g^{0.7}$, and no winds, we find:
\begin{equation}
z_{\rm crit} \approx \left\{ \begin{array}{ll}
			1.4 & ({\rm Normal~galaxies}, \Sigma_{\rm SFR} \la 0.02)\\
			5.8 \Sigma_{\rm SFR}^{0.23} - 1 & ({\rm Normal~galaxies}, \Sigma_{\rm SFR} \ga 0.02)\\
			5.7 \Sigma_{\rm SFR}^{0.23} - 1 & ({\rm Puffy~starbursts})\\
			7.4 \Sigma_{\rm SFR}^{0.23} - 1 & ({\rm Compact~starbursts}),
			\end{array} \right.
\end{equation}
where $\Sigma_{\rm SFR}$ is in units of $\Msun~\kpc^{-2}~\yr^{-1}$.  For the scenario with the B07 star-formation law, $B \propto \Sigma_g^{0.8}$, and winds in starbursts, we find:
\begin{equation}
z_{\rm crit} \approx \left\{ \begin{array}{ll}
			1.5 & ({\rm Normal~galaxies}, \Sigma_{\rm SFR} \la 0.02)\\
			6.6 \Sigma_{\rm SFR}^{0.24} - 1 & ({\rm Normal~galaxies}, \Sigma_{\rm SFR} \ga 0.02)\\
			4.8 \Sigma_{\rm SFR}^{0.06} - 1 & ({\rm Puffy~starbursts}, \Sigma_{\rm SFR} \la 0.2)\\
			6.5 \Sigma_{\rm SFR}^{0.24} - 1 & ({\rm Puffy~starbursts}, \Sigma_{\rm SFR} \ga 0.2)\\
			8.6 \Sigma_{\rm SFR}^{0.06} - 1 & ({\rm Compact~starbursts}, \Sigma_{\rm SFR} \la 0.8)\\
			9.0 \Sigma_{\rm SFR}^{0.24} - 1 & ({\rm Compact~starbursts}, \Sigma_{\rm SFR} \la 0.8).
			\end{array} \right.
\end{equation}
For the scenario with the B07 star-formation law, $B \propto \rho^{0.6}$, and winds in starbursts, we find:
\begin{equation}
z_{\rm crit} \approx \left\{ \begin{array}{ll}
			1.4 & ({\rm Normal~galaxies}, \Sigma_{\rm SFR} \la 0.02)\\
			5.6 \Sigma_{\rm SFR}^{0.21} - 1 & ({\rm Normal~galaxies}, \Sigma_{\rm SFR} \ga 0.02)\\
			4.5 \Sigma_{\rm SFR}^{0.04} - 1 & ({\rm Puffy~starbursts}, \Sigma_{\rm SFR} \la 0.2)\\
			5.8 \Sigma_{\rm SFR}^{0.21} - 1 & ({\rm Puffy~starbursts}, \Sigma_{\rm SFR} \ga 0.2)\\
			9.5 \Sigma_{\rm SFR}^{0.04} - 1 & ({\rm Compact~starbursts}, \Sigma_{\rm SFR} \la 0.7)\\
			10.0 \Sigma_{\rm SFR}^{0.21} - 1 & ({\rm Compact~starbursts}, \Sigma_{\rm SFR} \la 0.7).
			\end{array} \right.
\end{equation}
For the standard model of LTQ, with the K98 star-formation law, $B \propto \Sigma_g^{0.7}$, and no winds, we find:
\begin{equation}
z_{\rm crit} \approx \left\{ \begin{array}{ll}
			1.4 & ({\rm Normal~galaxies}, \Sigma_{\rm SFR} \la 0.02)\\
			6.6 \Sigma_{\rm SFR}^{0.25} - 1 & ({\rm Normal~galaxies}, \Sigma_{\rm SFR} \ga 0.02)\\
			6.8 \Sigma_{\rm SFR}^{0.25} - 1 & ({\rm Puffy~starbursts})\\
			10.2 \Sigma_{\rm SFR}^{0.25} - 1 & ({\rm Compact~starbursts}).
			\end{array} \right.
\end{equation}
Finally, for the model of LTQ with the K98 star-formation law, $B \propto \rho^{0.5}$, and winds in starbursts, we find:
\begin{equation}
z_{\rm crit} \approx \left\{ \begin{array}{ll}
			1.4 & ({\rm Normal~galaxies}, \Sigma_{\rm SFR} \la 0.02)\\
			5.6 \Sigma_{\rm SFR}^{0.23} - 1 & ({\rm Normal~galaxies}, \Sigma_{\rm SFR} \ga 0.02)\\
			4.5 \Sigma_{\rm SFR}^{0.04} - 1 & ({\rm Puffy~starbursts}, \Sigma_{\rm SFR} \la 0.3)\\
			5.5 \Sigma_{\rm SFR}^{0.23} - 1 & ({\rm Puffy~starbursts}, \Sigma_{\rm SFR} \ga 0.3)\\
			9.2 \Sigma_{\rm SFR}^{0.04} - 1 & ({\rm Compact~starbursts}, \Sigma_{\rm SFR} \la 0.7)\\
			9.9 \Sigma_{\rm SFR}^{0.23} - 1 & ({\rm Compact~starbursts}, \Sigma_{\rm SFR} \la 0.7).
			\end{array} \right.
\end{equation}

\begin{figure}
\centerline{\includegraphics[width=9cm]{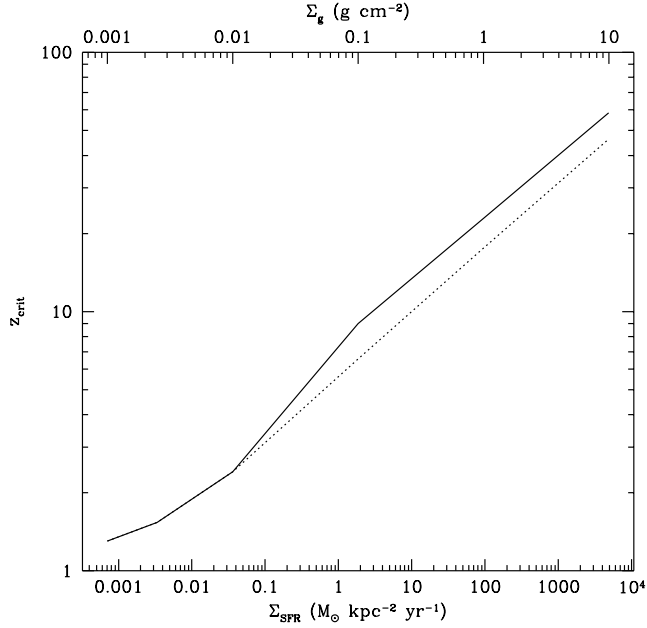}}
\figcaption[figure]{The redshift $z_{\rm crit}$ when the rest-frame 1.4 GHz synchrotron luminosity is suppressed by a factor of 3 from IC losses off the CMB, in our model with the B07 star-formation law, $B \propto \Sigma_g^{0.7}$, and no winds.  The dotted line is for puffy starbursts.  For starbursts, the suppression is only important at very high $z$, typically beyond those that will be observed by EVLA and SKA.\label{fig:zCrit}}
\end{figure}

\begin{deluxetable}{lllccccc}
\tablecaption{Model Properties}
\tablehead{\colhead{SFR} & \colhead{Winds} & \colhead{$B$} & \colhead{FRC Scatter (PS)\tablenotemark{a}} & \colhead{PS FRC Normalization ($\Delta q^{\prime}$)\tablenotemark{b}} & \multicolumn{3}{c}{Spectral slopes \tablenotemark{c}} \\  & & & & & Norm & CS & PS}
\startdata
K98 & N & $\Sigma_g^{0.7}$ & 1.7 (1.5) & 3.9 (-0.59) & 0.92 - 0.96 & 0.44 - 0.64 & 0.69 - 0.92\\
K98 & Y & $\rho^{0.5}$ & 1.7 (1.7) & 0.85 (0.07) & 0.88 - 0.94 & 0.43 - 0.61 & 0.78 - 0.85\\
B07 & N & $\Sigma_g^{0.7}$ & 1.7 (1.5) & 2.2 (-0.35) & 0.87 - 0.96 & 0.57 - 0.67 & 0.86 - 0.98\\
B07 & Y & $\Sigma_g^{0.8}$ & 2.2 (1.7) & 3.4 (-0.53) & 0.85 - 0.91 & 0.45 - 0.59 & 0.75 - 0.87\\
B07 & Y & $\rho^{0.6}$ & 2.0 (2.6) & 0.49 (0.31) & 0.87 - 0.91 & 0.49 - 0.64 & 0.88 - 0.91\\
\enddata
\tablenotetext{a}{Variation in the local $L^{\prime}_{\rm TIR}/L^{\prime}_{\rm radio}$ over normal galaxies and compact starbursts, as measured at $z = 0$ at $\nu^{\prime} = 1.4~\GHz$.  The value in parentheses is the variation in $L^{\prime}_{\rm TIR}/L^{\prime}_{\rm radio}$ for puffy starbursts alone at $z = 0$ with $\nu^{\prime} = 1.4~\GHz$.} 
\tablenotetext{b}{Average radio-brightness of puffy starbursts at $z = 0$, compared to the local normalization of the FRC for compact starbursts and normal galaxies.  $\Delta q^{\prime}$ is the offset in $q^{\prime}$ from its locally observed value for the puffy starbursts.}
\tablenotetext{c}{Range of instantaneous spectral slopes $\alpha$ at $\nu^{\prime} = 1.4~\GHz$ at $z = 0$.  Norm = normal galaxies; CS = compact starbursts; PS = puffy starbursts.}
\label{table:Models}
\end{deluxetable}


\begin{thebibliography}{}

\bibitem[Appleton et al.(2004)]{Appleton04} Appleton, P.~N., et al.\ 2004, \apjs~154, 147.

\bibitem[Bavouzet et al.(2008)]{Bavouzet08} Bavouzet, N., Dole, H., Le Floc'h, E., Caputi, K.~I., Lagache, G., \& Kochanek, C.~S.\ 2008, \aap, 479, 83.

\bibitem[Beck(2001)]{Beck01} Beck, R. 2001, \ssr, 99, 243.

\bibitem[Bell(2003)]{Bell03} Bell, E. F. 2003, \apj~586, 794.

\bibitem[Benn et al.(1993)]{Benn93} Benn, C.~R., Rowan-Robinson, M., McMahon, R.~G., Broadhurst, T.~J., \& Lawrence, A.\ 1993, \mnras, 263, 98 

\bibitem[Bernet et al.(2008)]{Bernet08} Bernet, M.~L., Miniati, F., Lilly, S.~J., Kronberg, P.~P., \& Dessauges-Zavadsky, M.\ 2008, \nat, 454, 302.

\bibitem[Beswick et al.(2008)]{Beswick08} Beswick, R.~J., Muxlow, T.~W.~B., Thrall, H., Richards, A.~M.~S., \& Garrington, S.~T.\ 2008, \mnras, 385, 1143 

\bibitem[Biggs \& Ivison(2008)]{Biggs08} Biggs, A.~D., \& Ivison, R.~J.\ 2008, \mnras~385, 893.

\bibitem[Bondi et al.(2007)]{Bondi07} Bondi, M., et al.\ 2007, \aap, 463, 519.

\bibitem[Bouch{\'e} et al.(2007)]{Bouche07} Bouch{\'e}, N., et al.\ 2007, \apj~671, 303.

\bibitem[Boyle et al.(2007)]{Boyle07} Boyle, B.~J., Cornwell, T.~J., Middelberg, E., Norris, R.~P., Appleton, P.~N., \& Smail, I.\ 2007, \mnras, 376, 1182 

\bibitem[Bressan et al.(2002)]{Bressan02} Bressan, A., Silva, L., \& Granato, G.~L.\ 2002, \aap, 392, 377.

\bibitem[Calzetti et al.(2000)]{Calzetti00} Calzetti, D., Armus, L., Bohlin, R.~C., Kinney, A.~L., Koornneef, J., \& Storchi-Bergmann, T.\ 2000, \apj~533, 682.

\bibitem[Carilli \& Yun(1999)]{Carilli99} Carilli, C.~L., \& Yun, M.~S.\ 1999, \apjl, 513, L13 

\bibitem[Carilli et al.(2008)]{Carilli08} Carilli, C.~L., et al.\ 2008, \apj, 689, 883.

\bibitem[Chapman et al.(2004)]{Chapman04} Chapman, S.~C., Smail, I., Windhorst, R., Muxlow, T., \& Ivison, R.~J.\ 2004, \apj~611, 732.

\bibitem[Chapman et al.(2005)]{Chapman05} Chapman, S.~C., Blain, A.~W., Smail, I., \& Ivison, R.~J.\ 2005, \apj, 622, 772 


\bibitem[Chary \& Elbaz(2001)]{Chary01} Chary, R., \& Elbaz, D.\ 2001, \apj, 556, 562 

\bibitem[Clemens et al.(2008)]{Clemens08} Clemens, M. S. et al. 2008, \aap~477, 95.

\bibitem[Condon(1989)]{Condon89} Condon, J.~J.\ 1989, \apj, 338, 13 

\bibitem[Condon et al.(1991)]{Condon91b} Condon, J.~J., Anderson, M.~L., \& Helou, G.\ 1991, \apj, 376, 95 

\bibitem[Condon et al.(1991)]{Condon91} Condon, J. J., Huang, Z.-P., Yin, Q. F., \& Thuan, T. X. 1991, \apj~378, 65.

\bibitem[Condon(1992)]{Condon92} Condon, J. J. 1992, \araa~30, 575.

\bibitem[Condon(2009)]{Condon09} Condon, J. J. 2009, Sensitive Continuum Survey with the SKA: Goals and Challenges (Manchester: SKA Program Development Office), SKA Memo \#114 (http://www.skatelescope.org/PDF/memos/114\_Memo\_Condon.pdf)

\bibitem[Coppin et al.(2006)]{Coppin06} Coppin, K., et al.\ 2006, \mnras, 372, 1621 

\bibitem[Cram(1998)]{Cram98} Cram, L.~E.\ 1998, \apjl, 506, L85.

\bibitem[Crutcher(1999)]{Crutcher99} Crutcher, R.~M.\ 1999, \apj, 520, 706 

\bibitem[Danese et al.(1987)]{Danese87} Danese, L., Franceschini, A., Toffolatti, L., \& de Zotti, G.\ 1987, \apjl, 318, L15 

\bibitem[de Jong et al.(1985)]{deJong85} de Jong, T., Klein, U., Wielebinski, R., \& Wunderlich, E. 1985, \aap~147, L6.

\bibitem[de Zotti et al.(2010)]{deZotti10} de Zotti, G., Massardi, M., Negrello, M., \& Wall, J.\ 2010, \aapr, 18, 1

\bibitem[Devlin et al.(2009)]{Devlin09} Devlin, M.~J., et al.\ 2009, \nat, 458, 737.

\bibitem[Dole et al.(2006)]{Dole06} Dole, H., et al.\ 2006, \aap, 451, 417

\bibitem[Downes \& Solomon(1998)]{Downes98} Downes, D., \& Solomon, P. M. 1998, \apj~507, 615.

\bibitem[Dwek \& Barker(2002)]{Dwek02} Dwek, E., \& Barker, M.~K.\ 2002, \apj, 575, 7 

\bibitem[Dye et al.(2007)]{Dye07} Dye, S., Eales, S.~A., Ashby, M.~L.~N., Huang, J.-S., Egami, E., Brodwin, M., Lilly, S., \& Webb, T.\ 2007, \mnras, 375, 725.


\bibitem[Fixsen et al.(2009)]{Fixsen09} Fixsen, D.~J., et al.\ 2009, arXiv:0901.0555.

\bibitem[Garn et al.(2009)]{Garn09} Garn, T., Green, D.~A., Riley, J.~M., \& Alexander, P.\ 2009, \mnras, 397, 1101 

\bibitem[Garn \& Alexander(2009)]{Garn09b} Garn, T., \& Alexander, P.\ 2009, \mnras, 394, 105 

\bibitem[Gavazzi et al.(1991)]{Gavazzi91} Gavazzi, G., Boselli, A., \& Kennicutt, R.\ 1991, \aj, 101, 1207.

\bibitem[Genzel et al.(2008)]{Genzel08} Genzel, R., et al.\ 2008, \apj, 687, 59.

\bibitem[Gervasi et al.(2008)]{Gervasi08} Gervasi, M., Tartari, A., Zannoni, M., Boella, G., \& Sironi, G.\ 2008, \apj, 682, 223 

\bibitem[Greve et al.(2009)]{Greve09} Greve, T.~R., Papadopoulos, P.~P., Gao, Y., \& Radford, S.~J.~E.\ 2009, \apj~692, 1432.

\bibitem[Haarsma \& Partridge(1998)]{Haarsma98} Haarsma, D.~B., \& Partridge, R.~B.\ 1998, \apjl, 503, L5 

\bibitem[Haarsma et al.(2000)]{Haarsma00} Haarsma, D.~B., Partridge, R.~B., Windhorst, R.~A., \& Richards, E.~A.\ 2000, \apj, 544, 641.

\bibitem[Hauser \& Dwek(2001)]{Hauser01} Hauser, M.~G., \& Dwek, E.\ 2001, \araa, 39, 249 

\bibitem[Helou et al.(1985)]{Helou85} Helou, G., Soifer, B. T., \& Rowan-Robinson, M. 1985, \apjl~298, 7.

\bibitem[Hughes et al.(2006)]{Hughes06} Hughes, A., Wong, T., Ekers, R., Staveley-Smith, L., Filipovic, M., Maddison, S., Fukui, Y., \& Mizuno, N.\ 2006, \mnras~370, 363.
	
\bibitem[Huynh et al.(2007)]{Huynh07} Huynh, M.~T., Jackson, C.~A., \& Norris, R.~P.\ 2007, \aj, 133, 1331.

\bibitem[Ibar et al.(2008)]{Ibar08} Ibar, E., et al.\ 2008, \mnras, 386, 953.

\bibitem[Ibar et al.(2010)]{Ibar10} Ibar, E., Ivison, R.~J., Best, P.~N., Coppin, K., Pope, A., Smail, I., \& Dunlop, J.~S.\ 2010, \mnras, 401, L53 

\bibitem[Iono et al.(2009)]{Iono09} Iono, D., et al.\ 2009, \apj, 695, 1537 

\bibitem[Ivison et al.(2010)]{Ivison10} Ivison, R.~J., et al.\ 2010, \mnras, 402, 245 

\bibitem[Jackson(2004)]{Jackson04} Jackson, C.~A.\ 2004, New Astronomy Review, 48, 1187 

\bibitem[Kennicutt(1998)]{Kennicutt98} Kennicutt, R. C. 1998, \apj~498, 541.

\bibitem[Kov{\'a}cs et al.(2006)]{Kovacs06} Kov{\'a}cs, A., Chapman, S.~C., Dowell, C.~D., Blain, A.~W., Ivison, R.~J., Smail, I., \& Phillips, T.~G.\ 2006, \apj, 650, 592 

\bibitem[Kronberg et al.(2008)]{Kronberg08} Kronberg, P.~P., Bernet, M.~L., Miniati, F., Lilly, S.~J., Short, M.~B., \& Higdon, D.~M.\ 2008, \apj~676, 70.

\bibitem[Kulsrud \& Zweibel(2008)]{Kulsrud08} Kulsrud, R.~M., \& Zweibel, E.~G.\ 2008, Reports on Progress in Physics, 71, 046901.

\bibitem[Lacki et al.(2009)]{Lacki09} Lacki, B. C., Thompson, T. A., \& Quataert, E. 2009, arXiv:0907.4161.

\bibitem[Lagache et al.(2005)]{Lagache05} Lagache, G., Puget, J.-L., \& Dole, H.\ 2005, \araa, 43, 727 

\bibitem[Law et al.(2009)]{Law09} Law, D.~R., Steidel, C.~C., Erb, D.~K., Larkin, J.~E., Pettini, M., Shapley, A.~E., \& Wright, S.~A.\ 2009, \apj, 697, 2057.

\bibitem[Le Floc'h et al.(2005)]{LeFloch05} Le Floc'h, E., et al.\ 2005, \apj, 632, 169 

\bibitem[Lisenfeld et al.(1996b)]{Lisenfeld96b} Lisenfeld, U., V\"olk, H. J., \& Xu, C. 1996, \aap~314, 745.

\bibitem[Loeb \& Waxman(2006)]{Loeb06} Loeb, A. \& Waxman, E. 2006, Journal of Cosmology and Astroparticle Physics 5,3.

\bibitem[Magnelli et al.(2009)]{Magnelli09} Magnelli, B., Elbaz, D., Chary, R.~R., Dickinson, M., Le Borgne, D., Frayer, D.~T., \& Willmer, C.~N.~A.\ 2009, \aap, 496, 57 

\bibitem[Micha{\l}owski et al.(2009)]{Michalowski09} Micha{\l}owski, M.~J., Hjorth, J., \& Watson, D.\ 2009, arXiv:0905.4499.

\bibitem[Micha{\l}owski et al.(2010)]{Michalowski10} Micha{\l}owski, M.~J., Watson, D., \& Hjorth, J.\ 2010, arXiv:1002.2636 

\bibitem[Miller \& Owen(2001)]{Miller01} Miller, N.~A., \& Owen, F.~N.\ 2001, \aj, 121, 1903.

\bibitem[Mobasher et al.(1999)]{Mobasher99} Mobasher, B., Cram, L., Georgakakis, A., \& Hopkins, A.\ 1999, \mnras, 308, 45.

\bibitem[Murphy et al.(2009b)]{Murphy09b} Murphy, E.~J., Kenney, J.~D.~P., Helou, G., Chung, A., \& Howell, J.~H.\ 2009, \apj, 694, 1435

\bibitem[Murphy et al.(2009a)]{Murphy09} Murphy, E.~J., Chary, R.-R., Alexander, D.~M., Dickinson, M., Magnelli, B., Morrison, G., Pope, A., \& Teplitz, H.~I.\ 2009, \apj, 698, 1380 

\bibitem[Murphy(2009)]{Murphy09c} Murphy, E.~J.\ 2009, \apj, 706, 482 

\bibitem[Pascale et al.(2009)]{Pascale09} Pascale, E., et al.\ 2009, arXiv:0904.1206 

\bibitem[Persic \& Rephaeli(2007)]{Persic07} Persic, M., \& Rephaeli, Y.\ 2007, \aap, 463, 481 

\bibitem[Pope et al.(2006)]{Pope06} Pope, A., et al.\ 2006, \mnras, 370, 1185.

\bibitem[Pope et al.(2008)]{Pope08} Pope, A., et al.\ 2008, \apj, 675, 1171.

\bibitem[Rengarajan(2005)]{Rengarajan05} Rengarajan, T.~N.\ 2005, Proc. 29th Int. Cosmic Ray Conf. (Pune), 3.

\bibitem[Rieke et al.(2009)]{Rieke09} Rieke, G.~H., Alonso-Herrero, A., Weiner, B.~J., P{\'e}rez-Gonz{\'a}lez, P.~G., Blaylock, M., Donley, J.~L., \& Marcillac, D.\ 2009, \apj, 692, 556.

\bibitem[Robinshaw et al.(2008)]{Robinshaw08} Robinshaw, T., Quataert, E., Heiles, C. 2008, \apj~680, 981.

\bibitem[Sajina et al.(2008)]{Sajina08} Sajina, A., et al.\ 2008, \apj, 683, 659.

\bibitem[Sargent et al.(2010)]{Sargent10} Sargent, M.~T., et al.\ 2010, \apjs, 186, 341 

\bibitem[Schmidt(1959)]{Schmidt59} Schmidt, M. 1959, \apj~129, 243.

\bibitem[Seiffert et al.(2009)]{Seiffert09} Seiffert, M., et al.\ 2009, arXiv:0901.0559.

\bibitem[Seymour et al.(2008)]{Seymour08} Seymour, N., et al.\ 2008, \mnras, 386, 1695.

\bibitem[Seymour et al.(2009)]{Seymour09} Seymour, N., Huynh, M., Dwelly, T., Symeonidis, M., Hopkins, A., McHardy, I.~M., Page, M., \& Rieke, G.\ 2009, arXiv:0906.1817 

\bibitem[Singal et al.(2009)]{Singal09} Singal, J., Stawarz, L., Lawrence, A., \& Petrosian, V.\ 2009, arXiv:0909.1997 

\bibitem[Solomon et al.(1997)]{Solomon97} Solomon, P.~M., Downes, D., Radford, S.~J.~E., \& Barrett, J.~W.\ 1997, \apj, 478, 144

\bibitem[Tacconi et al.(2006)]{Tacconi06} Tacconi, L.~J., et al.\ 2006, \apj, 640, 228

\bibitem[Thompson et al.(2006)]{Thompson06} Thompson, T. A. et al. 2006, \apj~645, 186.

\bibitem[Thompson, Quataert, \& Waxman(2007)]{Thompson07} Thompson, T. A., Quataert, E., Waxman, E. 2007, \apj~654, 219.

\bibitem[Thompson(2008)]{Thompson08} Thompson, T.~A.\ 2008, \apj~684, 212.

\bibitem[Thompson et al.(2009)]{Thompson09} Thompson, T.~A., Quataert, E., \& Murray, N.\ 2009, \mnras, 928.

\bibitem[Valiante et al.(2007)]{Valiante07} Valiante, E., Lutz, D., Sturm, E., Genzel, R., Tacconi, L.~J., Lehnert, M.~D., \& Baker, A.~J.\ 2007, \apj, 660, 1060.

\bibitem[van der Kruit(1971)]{vanDerKruit71} van der Kruit, P. C. 1971, \aap~15, 110.

\bibitem[van der Kruit(1973)]{vanDerKruit73} van der Kruit, P. C. 1973, \aap~29, 263.

\bibitem[Vlahakis et al.(2007)]{Vlahakis07} Vlahakis, C., Eales, S., \& Dunne, L. 2007, \mnras~379, 1042.

\bibitem[V\"olk(1989)]{Volk89} V\"olk, H. J. 1989, \aap~218, 67.

\bibitem[Walter et al.(2009)]{Walter09} Walter, F., Riechers, D., Cox, P., Neri, R., Carilli, C., Bertoldi, F., Weiss, A., Maiolino, R. 2009, \nat~457, 699.

\bibitem[Wang et al.(2004)]{Wang04} Wang, W.-H., Cowie, L.~L., \& Barger, A.~J.\ 2004, \apj, 613, 655 

\bibitem[Watabe et al.(2009)]{Watabe09} Watabe, Y., Risaliti, G., Salvati, M., Nardini, E., Sani, E., \& Marconi, A.\ 2009, \mnras, 396, L1.

\bibitem[Widrow(2002)]{Widrow02} Widrow, L.~M.\ 2002, Reviews of Modern Physics, 74, 775.

\bibitem[Wolfe et al.(2008)]{Wolfe08} Wolfe, A.~M., Jorgenson, R.~A., Robishaw, T., Heiles, C., \& Prochaska, J.~X.\ 2008, \nat, 455, 638.

\bibitem[Yang et al.(2007)]{Yang07} Yang, M., Greve, T.~R., Dowell, C.~D., \& Borys, C.\ 2007, \apj, 660, 1198.

\bibitem[Younger et al.(2008)]{Younger08} Younger, J.~D., et al.\ 2008, \apj~688, 59.

\bibitem[Younger et al.(2009)]{Younger09} Younger, J.~D., et al.\ 2009, \mnras, 394, 1685.

\bibitem[Younger et al.(2010)]{Younger10} Younger, J.~D., et al.\ 2010, arXiv:1003.4264 

\bibitem[Yun et al.(2001)]{Yun01} Yun, M. S., Reddy, N. A., \& Condon, J. J. 2001, \apj~554, 803.

\end{thebibliography}
\end{document}